\documentclass{aa}  
\usepackage{caption}
\usepackage{graphicx}
\usepackage{natbib}
\usepackage{lscape}
\usepackage{rotating}
\bibpunct{(}{)}{;}{a}{}{,} % to follow the A&A style
\usepackage{txfonts}
\usepackage{lscape}
\usepackage{breqn}
\usepackage{url}
\usepackage{longtable}
\usepackage{hyperref}
\def\ms{\hbox{m\,s$^{-1}$}}         %m.s -1
\def\kms{\hbox{km\,s$^{-1}$}}       %km.s -1
\begin{document}
\title{HADES RV Programme with HARPS-N at TNG\thanks{Based on observations made with the Italian Telescopio Nazionale Galileo (TNG), operated on the island of La Palma by the INAF - Fundaci\'on Galileo Galilei at the Roque de los Muchachos Observatory of the Instituto de Astrof\'isica de Canarias (IAC); photometric observations from the APACHE array located at the Astronomical Observatory of the Aosta Valley; photometric observations made with the robotic APT2 (within the EXORAP program) located at Serra La Nave on Mt. Etna.}}
\subtitle{VI. GJ\,3942\,b behind dominant activity signals}
\author{M. Perger\inst{1}
	\and I. Ribas\inst{1}
	\and M. Damasso\inst{2}    
	\and J.C. Morales\inst{1}
	\and L. Affer\inst{3}
	\and A. Suárez Mascareño\inst{4,5}
	\and G. Micela\inst{3}
	\and J. Maldonado\inst{3}	
	\and J. I. González Hernández\inst{4,6}
	\and R. Rebolo\inst{4,6,7}
	\and G. Scandariato\inst{8}   
	\and G. Leto\inst{8}		     
    \and R. Zanmar Sanchez\inst{8}        
	% ALPHABETICAL
	\and S. Benatti\inst{9}
	\and A. Bignamini\inst{10}
	\and F. Borsa\inst{11}
	\and A. Carbognani\inst{12}
	\and R. Claudi\inst{9} 
	\and S. Desidera\inst{9} 
	\and M. Esposito\inst{13} 
	\and M. Lafarga\inst{1}
	\and A. F. Martinez Fiorenzano\inst{14} 
	\and E. Herrero\inst{1}
	\and E. Molinari\inst{14,15}  
	\and V. Nascimbeni\inst{9,16}
	\and I. Pagano\inst{8} 		    
	\and M. Pedani\inst{14}   
	\and E. Poretti\inst{11,17,18}  
	\and M. Rainer\inst{11} 
	\and A. Rosich\inst{1}
	\and A. Sozzetti\inst{2}
	\and B. Toledo-Padr\'on\inst{4,6}	
}
\offprints{M. Perger, \email{perger@ice.cat}}
\institute{\inst{1}Institut de Ci\'encies de l'Espai (IEEC-CSIC), Campus UAB, Carrer de Can Magrans s/n, 08193 Cerdanyola del Vall\'es, Spain\\
%	\email{perger@ice.cat}\\
	\inst{2}INAF - Osservatorio Astrofisico di Torino, via Osservatorio 20, 10025 Pino Torinese, Italy\\
	\inst{3}INAF - Osservatorio Astronomico di Palermo, Piazza del Parlamento 1, 90134 Palermo, Italy\\
	\inst{4}Institute de Astrofísica de Canarias (IAC), 38205 La Laguna, Tenerife, Spain\\
	\inst{5}Observatoire Astronomique de l’Université de Genéve, Versoix, Switzerland\\
	\inst{6}Universidad de La Laguna (ULL), Dpto. Astrofísica, 38206 La Laguna, Tenerife, Spain\\
	\inst{7}Consejo Superior de Investigaciones Científicas (CSIC), 28006 Madrid, Spain\\
	\inst{8}INAF - Osservatorio Astrofisico di Catania, via S. Sofia 78, 95123 Catania, Italy\\
	\inst{9}INAF - Osservatorio Astronomico di Padova, Vicolo dell'Osservatorio 5, 35122 Padova, Italy\\
	\inst{10}INAF - Osservatorio Astronomico di Trieste, via Tiepolo 11, 34143 Trieste, Italy\\
	\inst{11}INAF - Osservatorio Astronomico di Brera, Via E. Bianchi 46, 23807 Merate (LC), Italy\\
	\inst{12}Osservatorio Astronomico della Regione Autonoma Valle d’Aosta, Fraz. Lignan 39, 11020 Nus (Aosta), Italy\\
	\inst{13}INAF - Osservatorio Astronomico di Capodimonte, Via Moiariello, 16, 80131 Napoli, Italy\\
	\inst{14}Fundación Galileo Galilei - INAF, Rambla José Ana Fernandez Pérez 7, 38712 Breña Baja, Spain\\
	\inst{15}INAF - Osservatorio Astronomico di Cagliari, Via della Scienza 5, 09047 Selargius (CA), Italy\\
	\inst{16}Dipartimento di Fisica e Astronomia, G. Galilei, Università degli Studi di Padova, Vicolo dell'Osservatorio 3, 35122 Padova, Italy\\
	\inst{17}Universit\' e de Toulouse; UPS-OMP; IRAP; 31400 Toulouse, France\\
	\inst{18}CNRS; IRAP; 14, avenue Edouard Belin, 31400 Toulouse, France
	}
\date{Received: \today; accepted XXX}
\abstract
% context heading (optional)
{Short- to mid-term magnetic phenomena on the stellar surface of M-type stars cannot only resemble the effects of planets in radial velocity data, but also may hide them.}
% aims heading (mandatory)
{We analyze 145 spectroscopic HARPS-N observations of GJ\,3942 taken over the past five years and additional photometry to disentangle stellar activity effects from genuine Doppler signals as a result of the orbital motion of the star around the common barycenter with its planet.}
% methods heading (mandatory)
{To achieve this, we use the common methods of pre-whitening, and treat the correlated red noise by a first-order moving average term and by Gaussian-process regression following an MCMC analysis.}
% results heading (mandatory)
{We identify the rotational period of the star at 16.3\,days and discover a new super-Earth, GJ\,3942\,b, with an orbital period of 6.9\,days and a minimum mass of 7.1\,M$_{\oplus}$. An additional signal in the periodogram of the residuals is present but we cannot claim it to be related to a second planet with sufficient significance at this point. If confirmed, such planet candidate would have a minimum mass of 6.3\,M$_{\oplus}$ and a period of 10.4\,days, {\bf which might indicate a 3:2 mean-motion} resonance with the inner planet.}
% conclusions heading (optional), leave it empty if necessary
{ }
\maketitle
%________________________________________________________________
%
\section{Introduction} \index{in} \label{in}
In spectroscopic time-series data we can measure the radial velocity (RV) variations of a star as a result of the Doppler shift. Orbiting planets induce periodic signals originating from the wobble of the star around their common barycenter. Other sources of RV variations are related to different kinds of surface phenomena induced by short-term effects like stellar oscillations and granulation \citep{2015A&A...583A.118M, 2011A&A...525A.140D} and the magnetic activity of the host star. This includes mid-term effects like dark spots and bright faculae moving with the stellar rotation \citep{2016MNRAS.457.3637H, 2016A&A...586A.131H, 2017MNRAS.468.4772S}. And it includes long-term effects connected with the location or number of surface active regions and the magnetic cycle of the star \citep{2016A&A...587A.103L, 2010A&A...511A..54S}. The changes in RV result most importantly from the different amount of photons coming from the blue-shifted and from the red-shifted limbs of the rotating star, and also from the effects of convective blue shift and its suppression in magnetically active regions \citep{2016MNRAS.457.3637H, 1982Natur.297..208L}.

In recent years exoplanet surveys are paying considerable attention to M dwarfs since they provide the possibility to detect, characterize and understand Earth-like planets. For Sun-like stars, the current technical RV limit of $\sim$1\,\ms is still an order of magnitude too large to accomplish this goal. In addition to the advantage of their larger RV signals, M dwarfs have closer habitable zones with shorter orbital periods and higher transit probabilities. However different surface phenomena rotating approximately with the stellar rotation period induce signals into the RV data that can mimic planets as seen in stars like GJ\,581 \citep{2014ApJ...788..160J, 2014Sci...345..440R}, GJ\,667C \citep{2013A&A...556A.126A} or GJ\,674 \citep{2007A&A...474..293B}, but also hide planets of Earth-mass \citep{2014ApJ...794...51H}.

The exoplanetary community is investing significant effort into being able to disentangle magnetic activity and planetary signals and into understanding and correcting for the influences of magnetic phenomena on the RVs. Those include computational efforts to improve the search for periodicities in RV data \citep[see e.g.,][for an overview]{2016A&A...593A...5D, 2017A&A...598A.133D}, the construction of instruments observing at near infrared wavelengths like CARMENES \citep{2014SPIE.9147E..1FQ}, GIARPS \citep{2016SPIE.9908E..1AC}, SPIRou \citep{2014SPIE.9147E..15A}, or HPF \citep{2012SPIE.8446E..1SM}, and the conduction of spectroscopic surveys tailored to M dwarfs \citep{2013A&A...549A.109B, 2013A&A...549A..48T, 2015AAS...22525801I, 2015A&A...577A.128A, 2016Natur.536..437A}.

With our "HArps-n red Dwarf Exoplanet Survey" (HADES) we aim at the goal of detecting rocky planets by monitoring 78 bright M0 to M3 stars. This is a coordinated effort of Spanish (IEEC-CSIC, IAC) and Italian (INAF) institutions. The observations began in August 2012 and, as of March 2017, we have obtained approximately 3\,700 spectra. We have already detected two low-mass planets around GJ\,3998 \citep{2016A&A...593A.117A} and a potentially habitable one around GJ\,625 \citep{2017arXiv170506537S}. And we have also described in detail the stellar rotation characteristics of our sample by analyzing their activity indices \citep{2017A&A...598A..27M, 2017A&A...598A..28S}. Additionally, we made simulations to predict the outcome of our survey finding an average underlying non-correlated activity jitter of 2.3\,\ms \citep{2017A&A...598A..26P}.

In this work we search for planetary companions around GJ\,3942 (HIP~79126) by analyzing its RV time-series measurements obtained with the HARPS-N spectrograph. GJ\,3942 is a low-mass star with spectral type M0, located in the constellation of Draco and with kinematics suggesting membership to the young disc population \citep{2017A&A...598A..27M}. The analysis of its spectroscopic properties \citep{2017A&A...598A..27M, 2012AJ....143...93R, 2008A&A...478..507M} reveals solar metallicity, intermediate rotation and average-to-weak magnetic activity. Just recently the star was labeled as new candidate RV-variable star from 7 RV data points in the near-infrared obtained by \cite{2016ApJ...822...40G}. We provide an overview of the basic properties of GJ\,3942 in Table~\ref{T2}.

\begin{table}[htb]
	\caption{Stellar parameters of GJ\,3942}  \label{T2}
	\small
	\centering
	\begin{tabular}{lll}
		\hline \hline
		\noalign{\smallskip}
		R.A. &  16h:09m:3.14s & (1) \\
		Dec.  & +52$^{\circ}$:56':37.9'' & (1)  \\  \hline
		Spectral type & M0 & (2) \\
		Distance & 16.93$\pm$0.30~pc   & (1) \\
		$M$ & 0.63$\pm$0.07~$M_{\odot}$  & (2)\\
		$R$ & 0.61$\pm$0.06~$R_{\odot}$  & (2)\\
		$T_{\rm eff}$  & 3867$\pm$69~K  & (2)\\
		$\log g$ & 4.65$\pm$0.06  & (2)\\
		$[Fe/H]$ & $-$0.04$\pm$0.09  & (2)\\
		$\log R'_{HK}$ & 4.55$\pm$0.05 & (3) \\ \hline
		$\log L/L_{\odot}$  & $-$1.12$\pm$0.10  & (2) \\
		$\log L_{\rm X}$  & 27.4\,erg\,s$^{-1}$  & (4) \\
		$B$ & 11.59$\pm$0.10~mag & (5) \\
		$V$ & 10.25$\pm$0.04~mag & (5) \\
		$R$ & 9.4~mag &  (6) \\
		$I$ & 8.7~mag &  (7) \\
		$J$ & 7.185$\pm$0.020~mag & (8)\\
		$H$ & 6.525$\pm$0.020~mag & (8) \\
		$K$ & 6.331$\pm$0.018~mag & (8) \\    \hline
		$\pi$ & 58.95$\pm$0.25~mas & (9) \\
		$\mu_{\alpha}$ & $+$204.16$\pm$0.09~mas~yr$^{-1}$ & (9)\\
		$\mu_{\delta}$  & $+$62.00$\pm$0.10~mas~yr$^{-1}$ & (9) \\
		$dv/dt$  & 1.8$\pm$1.5~cm~s$^{-1}$~yr$^{-1}$ & (10) \\ 
		$v \sin i$ & 1.67$\pm$0.30~km s$^{-1}$ & (2)\\
		U & $-$0.47~km s$^{-1}$ & (2)\\
		V & $-$0.89~km s$^{-1}$ & (2)\\
		W & $-$25.31~km s$^{-1}$ & (2)\\	 \hline
		Inner HZ & 0.29~AU, 73~days & (11) \\
		Outer HZ & 0.55~AU, 187~days & (11) \\
		\noalign{\smallskip}
		\hline \hline
	\end{tabular}
	\tablefoot{\tablefoottext{1}{\cite{2007A&A...474..653V}}, \tablefoottext{2}{\cite{2017A&A...598A..27M}},
		\tablefoottext{3}{following \cite{2015MNRAS.452.2745S}},\tablefoottext{4}{\cite{1999A&AS..135..319H}}, \tablefoottext{5}{\cite{2000A&A...355L..27H}}, \tablefoottext{6}{\cite{2012yCat.1322....0Z}}, \tablefoottext{7}{\cite{2003AJ....125..984M}}, \tablefoottext{8}{\cite{2003tmc..book.....C}}, \tablefoottext{9}{\cite{2016A&A...595A...2G}}, \tablefoottext{10}{secular acceleration following \cite{2014ApJ...781...28M}}, \tablefoottext{11}{habitable zones (HZ) following \cite{2013ApJ...765..131K}}}
\end{table}

In Sect.\,\ref{de} we present the spectroscopic and photometric observations obtained and give a first overview of the periodic signals which are present in the data. In Sect.\,\ref{da} we explain the search for planets in detail and use various state-of-the-art approaches to correct for the effects of magnetic activity. This leads to the detection of a super-Earth sized planet GJ\,3942\,b and a second planetary candidate that we discuss in Sect.\,\ref{di}. The conclusions of our work are given in Sect.\,\ref{con}.

\section{Data reduction and preliminary analysis} \index{de} \label{de}

We obtained optical spectra with the Northern ``High Accuracy Radial velocity Planet Searcher'' \citep[HARPS-N,][]{2012SPIE.8446E..1VC}, connected by fibers to the Nasmyth B focus through a Front End Unit of the 3.58-m Telescopio Nazionale Galileo (TNG) in La Palma, Spain. It is a fiber-fed, cross-dispersed echelle spectrograph with a spectral resolution of 115\,000, covering a wavelength range from 3\,830 to 6\,900~$\AA$. We observed with fixed integration times of 900~s, to obtain data of sufficient signal-to-noise ratio (S/N$>$20) and to average out potential short-term periodic oscillations of the star \citep{2011A&A...525A.140D}, although they do not seem to be present in M dwarfs \citep{2016MNRAS.459.3551B}.

We reduced the raw data with the YABI tool \citep{Hunter_researchopen, 2015A&A...578A..64B}, which implements the DRS data reduction pipeline. It uses the classical optimal extraction method by \cite{1986PASP...98..609H} and includes bias and background subtraction and flat-fielding to deliver cosmic ray-corrected, wavelength-calibrated spectra. Furthermore, it calculates the cross-correlation function (CCF), which is the correlation of a spectrum with a M2-type template mask in velocity space. Various parameters, including the RV, are calculated from this CCF. For more details see \cite{2012SPIE.8446E..1VC} or the DRS manual\footnote{\href{url}{http://www.tng.iac.es/instruments/harps/data/\\
HARPS-N\_DRSUserManual\_1.1.pdf}}. In our case, we decided not to use RVs determined by the DRS but rather use the Java-based ``Template-Enhanced Radial velocity Re-analysis Application'' \cite[TERRA,][]{2012ApJS..200...15A}. It handles the full process of unpacking the HARPS-N archive files in the DRS output. The template-matching algorithm has been shown to deliver more accurate RVs in the case of M-type stars \citep{2017A&A...598A..26P}. 

GJ\,3942 was observed on 145 occasions resulting in spectra with average S/N of 56.2. The full time series dataset spanning 1\,203 nights or 3.3 years is provided in Table~\ref{Table}. As a quality check, we searched for correlations between the RVs and the S/N (which would be indicative of charge-transfer inefficiency effects), for obvious flare events (seen as emission in lines that have a contribution from the stellar chromosphere) or other irregularities and did a 5$\sigma$-clipping on the RV measurements. As a result, we rejected 3 data points with S/N$<$20 from further analysis. With the 142 remaining spectra, we determine an RV scatter and mean error of 6.01~\ms and 1.13~\ms, respectively. For comparison, these values are 6.56~\ms and 1.92~\ms in the case of the DRS. We also obtain an absolute average RV value of 18.7047$\pm$0.0066~\kms.

\subsection{Instrumental radial velocity drift} \index{de:ins} \label{de:ins}

\begin{figure}[tb]
	\resizebox{\hsize}{!}{\includegraphics[clip=true]{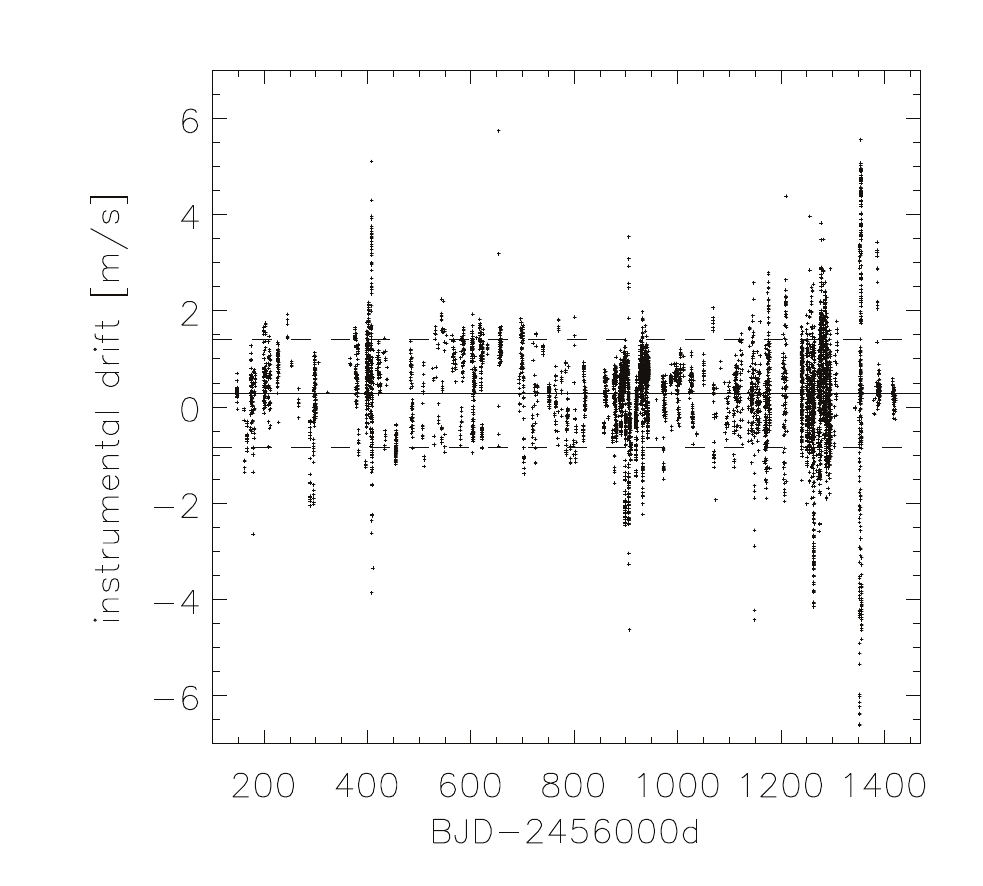}}
	\caption{All available 8\,199 instrumental drifts distributed over 356 nights. The solid line indicates the mean value and the dashed lines the 1$\sigma$-interval.}
	\label{F1}
\end{figure}

The HARPS-N instrument implements the possibility of using calibration light simultaneously with the target to measure the instrumental drift of the RV during science observations. A second fiber can collect light from a stable hollow-cathode lamp of ThAr, resulting in a simultaneous wavelength solution. Until August 2015, we did not use such calibration strategy because for t$_{exp}>$200~s it was supposed to contaminate the RV measurements from the science fiber. As we learned during our program, this does not seem to be the case at least for the brighter objects ($V<$10.5~mag), like GJ\,3942. Therefore, of the 142 measurements, only 41 have simultaneous calibration from which instrumental drift values can be determined. However, measurements taken by the Italian GAPS \citep[Global Architecture of Planetary Systems,][]{2013A&A...554A..28C, 2013A&A...554A..29D} project did include simultaneous calibration and this provides the opportunity of estimating instrumental drift corrections for a total of 51 additional nights on which GJ\,3942 was observed.

To estimate drift corrections we make use of 8\,199 measurements available from Spanish and Italian observations. They are distributed over 356 nights with an average of 23.03 measurements per night and are shown in Fig.~\ref{F1}. In some nights, the RV drift varied by as much as 12\,\ms. We want to use the data of a specific night to estimate the instrumental drift at the time of the observation of GJ\,3942. To evaluate the best method, we randomly take out one measured value and calculate the predicted value from the remaining measurements from the given night using several approaches. We show the mean values and rms of the differences in Table\,\ref{T1} using 100 trials. For the 51 nights covered by instrumental drift values, we find that the best results are obtained by averaging the available measurements in a time interval of $\pm$1\,h. Thus, we apply such methodology (method 4 in Table\,\ref{T1}) and add quadratically an uncertainty of 0.466\,\ms to each measurement. For those 50 measurements where no RV drift calculation is possible we consider an additional uncertainty of 1.025\,\ms added in quadrature, which corresponds to the total rms of all drift values. The instrumental drifts for each observation are provided in Table\,\ref{Table}.

\begin{table}[htb]
	\caption{Different methods to estimate instrumental drifts and the resulting rms of the data. The value in italics is the one used in this work.} \label{T1}
	\centering
	\small
	\begin{tabular}{l|lc}
		\hline \hline
		method & data treatment & mean error [m s$^{-1}$] \\ \hline
		\noalign{\smallskip}
		1) & mean value of night & 0.576$\pm$0.049 \\
		2) & linear fit on night & 0.530$\pm$0.079  \\
		3) & average of 2 closest neighbors & 0.482$\pm$0.081  \\
		\textit{4)} & \textit{average of measurements $\pm$1~h} &\textit{0.466$\pm$0.048}  \\
		5) & average of measurements $\pm$2~h & 0.471$\pm$0.050  \\
		6) & average of measurements $\pm$3~h & 0.485$\pm$0.053 \\
		7) & average of measurements $\pm$4~h & 0.491$\pm$0.052  \\
	\noalign{\smallskip}
			\hline \hline
	\end{tabular}
\end{table}

\subsection{Activity indices} \index{de:act} \label{de:act}

To be able to distinguish signals induced by activity from genuine planetary Doppler signals, we calculate activity indices of magnetically-sensitive features, which are primarily associated to emission from the stellar chromosphere. These are the central cores of the \ion{Ca}{II} H \& K lines and the H$\alpha$ line \citep{2011A&A...534A..30G}.

The $S$ index measures the relative flux of the \ion{Ca}{II} H \& K emission lines compared to a local continuum. The lines are formed in the hot plasma of the chromospheres of stars, hence vary with the strength of the stellar magnetic field \citep{2012ApJS..200...15A}, and measure the lower chromosphere \citep{2011A&A...534A..30G}. We calculate the $S$ index with the definitions by \cite{1991ApJS...76..383D}. This consists of adding the total flux in the central cores of the two lines and calculating the ratio to the sum of fluxes corresponding to two triangular windows on either side of the lines: $S=\alpha \cdot(F_{\rm H}+F_{\rm K})/(F_{\rm R}+F_{\rm V})$. We further modify the wavelength windows following \cite{2011A&A...534A..30G} to avoid adding information from the undesired photospheric contribution in the wings of the line, slightly shift the continuum windows and use a value $\alpha$ of 18.4 following \cite{2011arXiv1107.5325L}. This method produces values that are not in the same scale as the Mount Wilson index \citep{1991ApJS...76..383D, 2015MNRAS.452.2745S}, but we are more interested in defining a precise indicator that is useful to search for small periodic variations. The H$\alpha$ index is calculated as $H \alpha=F_{H \alpha}/(F_{R}+F_{V})$ \citep{2009A&A...495..959B}, modified by using a broader central window to include more contribution from the chromosphere following \cite{2011A&A...534A..30G}, and slightly different continuum windows.

Uncertainties for the indices are calculated as Gaussian errors using Poisson noise on the spectral data. The $S$ and H$\alpha$ indices show relative errors of 1.8\% and 1.6\%, respectively. The index calculated from the \ion{Na}{I} doublet \citep{2007MNRAS.378.1007D} was not used in the present study because our preliminary analysis did not show it to be a good activity proxy for this target.

The DRS pipeline and the CCF technique also provide us with additional quantities. Magnetic phenomena on the stellar surface affect the shape of the lines and this is imprinted on the CCF. Variations of the parameters that describe the CCF therefore may be correlated with activity. Three of such parameters are calculated automatically by the pipeline: the Bisector Inverse Slope \citep[BIS,][]{1989PASP..101..832G}, the full width at half maximum (FWHM), and the peak value (contrast) of the Gaussian fit to the CCF. But the shape of the CCF of our target shows two prominent side lobes caused by the high density of spectroscopic features and is not well fitted by a Gaussian \citep[e.g.,][]{2016A&A...593A.117A, 2017A&A...597A.108S}. This explains the higher dispersion of the DRS RV values and we therefore prefer not to use the CCF indices as activity proxys in our analysis.

\subsection{Signal identification} \index{de:sig} \label{de:sig}

The classical approach to find periodic signals in unevenly-sampled time-series data is the Lomb-Scargle- \citep{1976Ap&SS..39..447L, 1982ApJ...263..835S} or the Generalized Lomb-Scargle (GLS) periodogram \citep{2009A&A...496..577Z}, which is a sinusoidal fit on error-weighted, shifted values using the function:
\begin{equation}
f(t) =c_{0} + c_{1} \cos \frac{2 \pi}{P} (t-t_{0}) + c_{2} \sin \frac{2 \pi}{P} (t-t_{0}). \label{For1}
\end{equation}
The parameters c$_{0}$, c$_{1}$, and c$_{2}$ are fitted to the data for each period ($P$) step, and a time shift $t_{0}$. In the case of a Keplerian orbit, this fit can be specified by
\begin{equation}
RV(t) = \gamma + K ( \cos(\omega + TA(t) ) + e \cos \omega ), \label{For2}
\end{equation}	
where $\gamma$ is the RV offset, $K$ the semi-amplitude of the signal, $e$ and $\omega$ the eccentricity and argument of periastron of the planetary orbit, respectively, and $TA$ the true anomaly of the planet \citep{2009ApJS..182..205W}.

As a measure of the significance of a possible signal, the False Alarm Probability (FAP) is calculated by bootstrapping data with 10\,000 permutations. Common practice indicates that a signal at the 1\% FAP level is defined as suggestive, while a 0.1\% FAP level signal is considered to be statistically significant. An analytical formula for the FAP is given by \cite{1986ApJ...302..757H}. We analyzed the RVs, the $S$ index, the H$\alpha$ index, and their respective residuals using the GLS.

\begin{figure}[tb]
	\resizebox{\hsize}{!}{\includegraphics[clip=true]{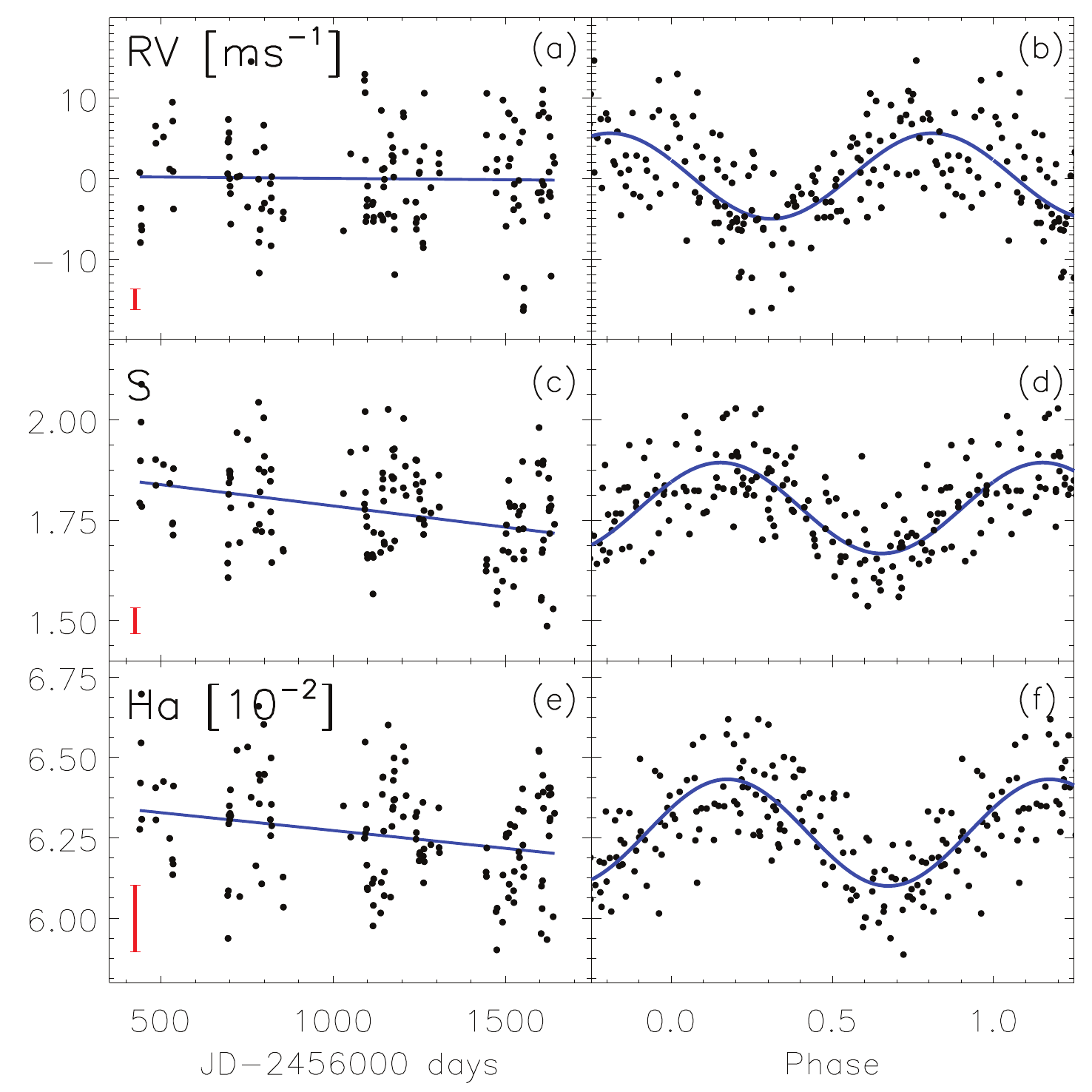}}
	\caption{Time series data of (from top to bottom panels) radial velocities (in m s$^{-1}$), $S$ index, and H$\alpha$ index (in 10$^{-2}$) are shown in panels (a), (c), (e), respectively, as observed (black dots) and we show their linear trend indicated by the blue lines. The mean errors of each data set are illustrated by the red line on the lower left side of each panel.
	Panels (b), (d), and (f) show the de-trended (i.e., linear fit subtracted) data folded to the most prominent periodicity at 16.3\,days and with zero phase corresponding to the first observation (black dots) and a simple sinusoidal fit (blue curve).}
	\label{F8a}
\end{figure}

The left panels of Fig.~\ref{F8a} show the different measurements corresponding to the RVs, $S$ index and H$\alpha$ index, where the four observing seasons are evident from the clustering. The global trends of the data yield negative slopes with modest significance. For the $S$, and $H_{\alpha}$ indices slopes of $-$0.039 and $-$0.040 per year are found, respectively, while the overall slope of the RVs is insignificant, at 0.12~m~s$^{-1}$ per year. We subtract those trends from the data for subsequent analysis to correct for any long-term effects resulting from a possible magnetic cycle \citep{2017A&A...598A.133D}. We note that a more detailed treatment of the trends leads to similar results. In the right panels of Fig.~\ref{F8a} we show the different de-trended datasets phase folded to the most prominent periodic signal at 16.27\,days (the mean of the best fits for the three values) and the best sinusoidal fits. The origin of phases is set to day one of our observations in all three cases. We observe a phase shift of approximately 120~deg between the activity indices and the RVs. This is not unusual and, e.g., \cite{2014A&A...566A..35S} found a shift of 15\,deg for their \ion{Ca}{II} index possibly connected to the location of activity phenomena on the stellar surface.

\begin{figure}[tb]
	\resizebox{\hsize}{!}{\includegraphics[clip=true]{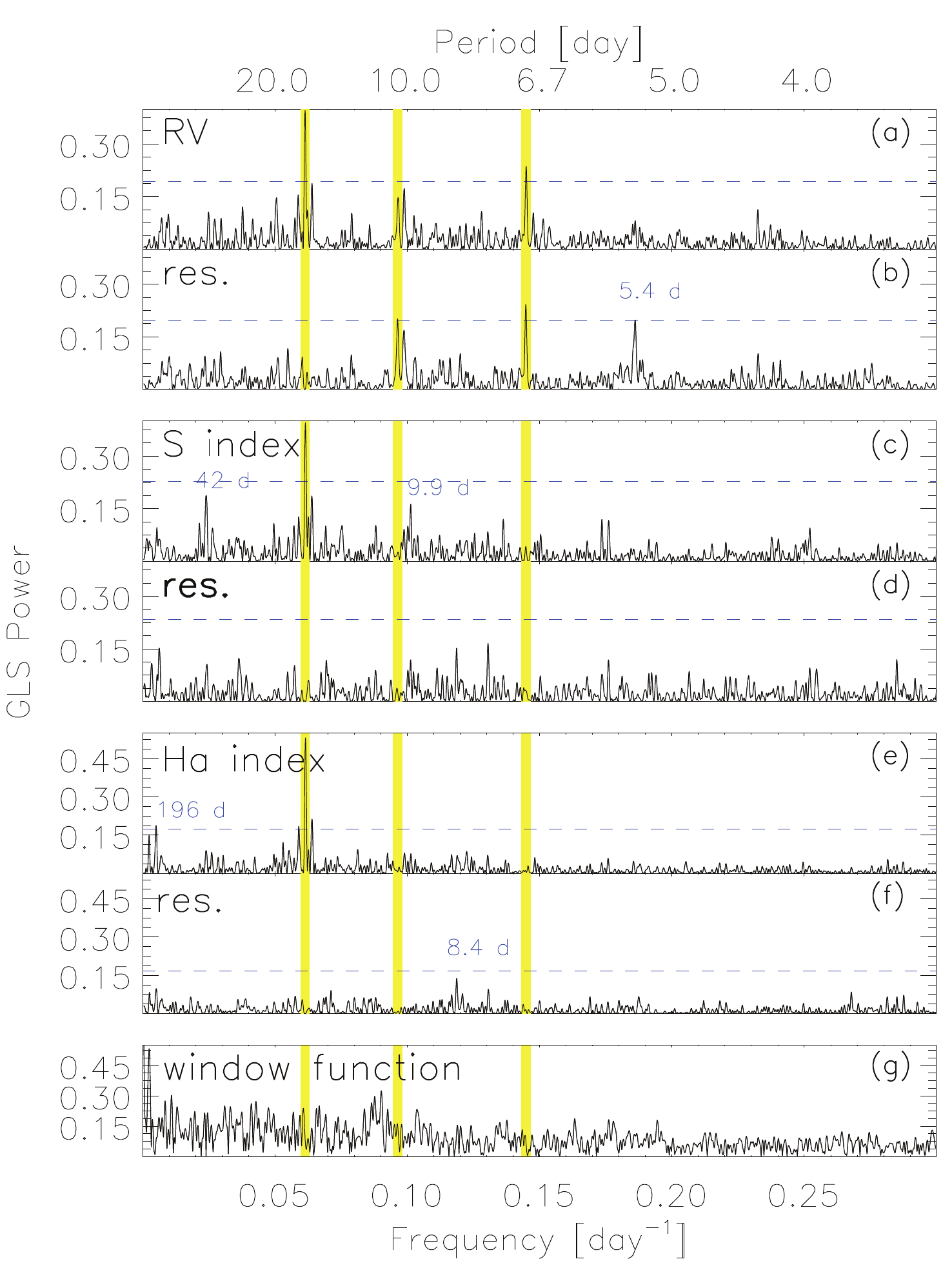}}
	\caption{Generalized Lomb-Scargle periodograms of the de-trended time-series data and of the residuals after correcting for the 16.3-day signal with a simple sinusoidal fit. In panels (a) and (b) this is shown for the radial velocities, in panel (c) and (d) for the $S$ index, and in panel (e) and (f) for the H$\alpha$ index. Additionally, we show in panel (g) the window function of the time series. Yellow lines indicate the important periodic signals for GJ\,3942, namely 16.3, 10.4, and 6.9\,days and we also show additional periodicities mentioned in the text. The dashed blue horizontal line shows the 1\% FAP level of the respective dataset.}
	\label{F8b}
\end{figure}

In Fig.~\ref{F8b} we show the GLS periodograms of the three de-trended time-series and their residuals from frequencies of 0.0 to 0.3\,day$^{-1}$ including all peaks clearly above noise level. In panel (a) we show the periodogram of the RVs with significant signals at 16.3 and 6.9\,days. Together with the signals at around 10\,days, those are the main periodicities in our data and are marked by the yellow bands in the different panels. We correct for the most prominent signal in the RV data at 16.29$\pm$0.22\,days by a simple sinusoid (see Fig.\,\ref{F8a}), analyze the residuals with the GLS, and show the periodogram in panel (b). The subtraction cleans up the periodogram significantly but preserves the 6.9 day peak, with a FAP below 0.1\%, as well as the two different peaks at 10.1 and 10.4\,days, which increase in significance to reach 1\% FAP. Additionally, at the same FAP level, a signal at 5.4\,days emerges. The $S$ index periodogram in panel (c) does not show any significant signal besides that at 16.3 days. Only some excess power is observed at approximately 9.9 and 42\,days, which disappear after removing the 16.26$\pm$0.22-day period as seen in the periodogram in panel (d). There, signals at $\>$100, 7.5, and 9\,days increase in power, with the latter two likely related to the first harmonic of 16.3\,days. The strong signal at 16.3\,days is also found in the H$\alpha$ index dataset in panel (e). Here, a suggestive periodicity of approximately 200 days is present as well. Removing the signal at 16.25$\pm$0.22\,days leaves only a peak connected to the first harmonic of the 16.3-day periodicity in the periodogram in panel (f). The window function (WF) as shown in panel (g) does not contribute to any of our important signals.

\begin{figure}[tb]
	\resizebox{\hsize}{!}{\includegraphics[clip=true]{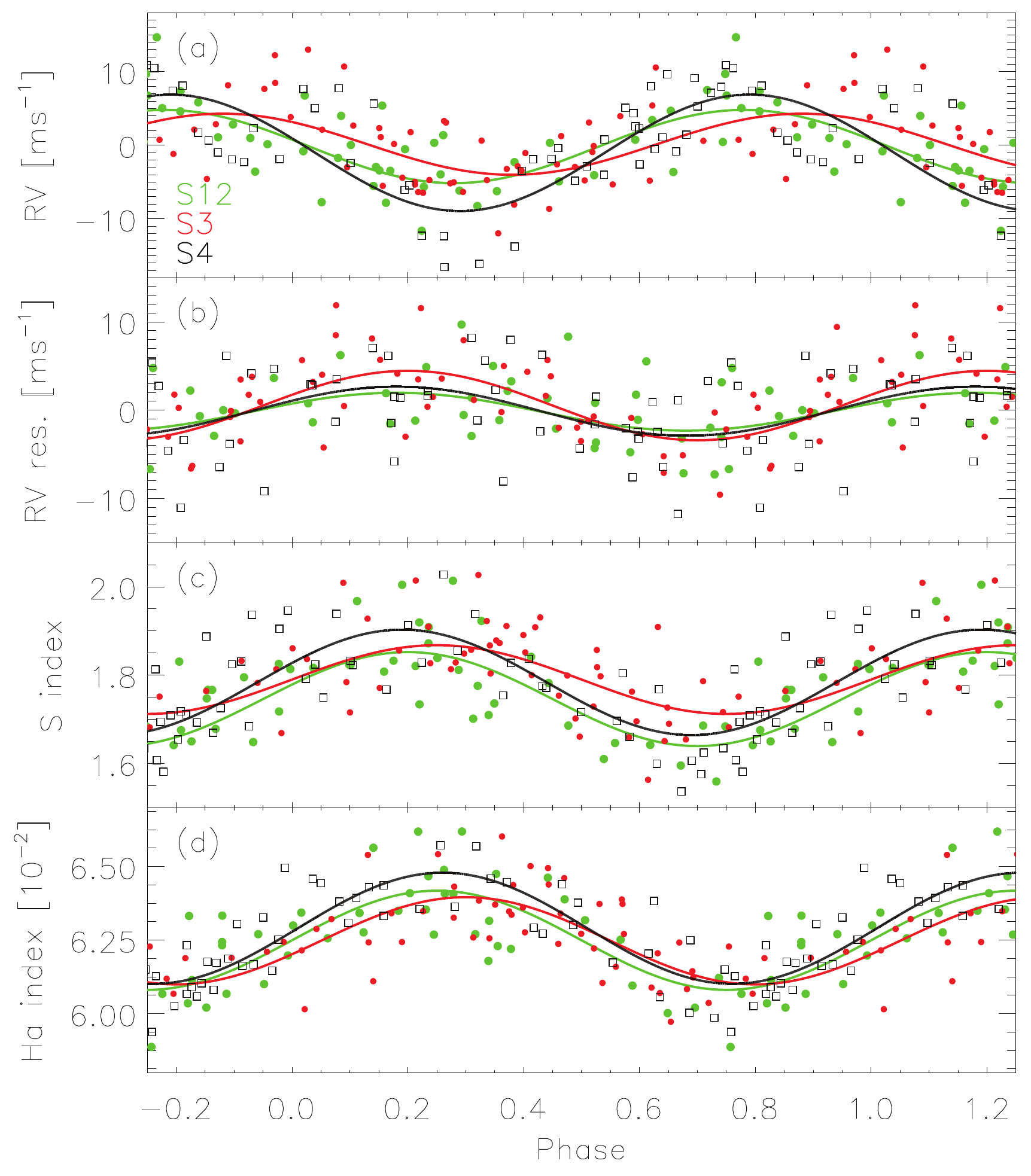}}
	\caption{Illustration of changing characteristics of signals in RV, RV residual, and activity index data of GJ\,3942. It is grouped into seasons 1 \& 2 (S12, green dots), season 3 (S3, red dots), and season 4 (black squares). The best purely sinusoidal fit for each dataset is shown. Panel (a) shows the RV data folded in phase with a 16.29-day period; panel (b) the residuals after correcting for the best sinusoidal 16.3-day period fit and phase folded with 6.91\,days; panel (c) the $S$ index data folded in phase with a 16.26-day period; and panel (d) the $H \alpha$ index data phase folded with 16.25\,days.}
	\label{F9}
\end{figure}

This most prominent periodicity in our RV data at 16.3\,days is clearly related to rotational modulation and may have suffered variations over the observational time span \citep{2014A&A...566A..35S, 2017A&A...598A.133D}. To investigate this, we performed sinusoidal fits to the RV data but grouping them into seasons. We considered seasons 1 and 2 (S12) together to increase the number of measurements in such a way that 43, 53, and 46 data points were used for S12 (green dots), S3 (red dots), and S4 (black squares), respectively. Fits with a constant period of 16.29\,days are shown in panel (a) of Fig.~\ref{F9}. In the diagram, we can observe a phase shift of S3 of approximately 35\,deg, while S4 shows an increase of the semi-amplitude of approximately 70\%. Fitting a free period shows as well its variability since for S3 alone the main signal is at a lower value of 15.9\,days. The same trend is also seen for the activity indices in panels (c), and (d).

The variation of the prominent signal at least on a yearly basis over the observed time span is another strong argument for the signal to be induced by activity phenomena present on the stellar surface and rotating with the star \citep{2016A&A...587A.103L, 2016A&A...585A.134D,2013A&A...551A.101M}. The changing characteristics of the fit are likely explained by the changing locations, sizes, and strengths of the magnetically-active regions and by differential rotation. The periodogram peak at 6.9\,days is not affected by the subtraction of the 16.3-day signal and does not show significant power in any of the activity indices. We therefore conclude that this RV periodic signal must be a result of the Keplerian motion of an orbiting planet, GJ\,3942\,b. In panel (b) of Fig.\,\ref{F9}, where the RV residuals are shown, no phase shift in S3 is visible, but an increase of the signals amplitude, whereas the parameters in general seem to be more stable than for the 16.3-day period. While this planetary signal is quite robust, more complicated is the interpretation of the signals between 7.5 and 10.5\,days. In this interval, the periodicity at approximately 8.2\,days corresponds to the first harmonic of the activity signal \citep{2011A&A...528A...4B}. The peaks around 10.4\,days, which are not highly significant, do not have a straightforward interpretation and the different possibilities are discussed in Sect.\,\ref{da}.

\begin{figure}[tb]
	\resizebox{\hsize}{!}{\includegraphics[clip=true]{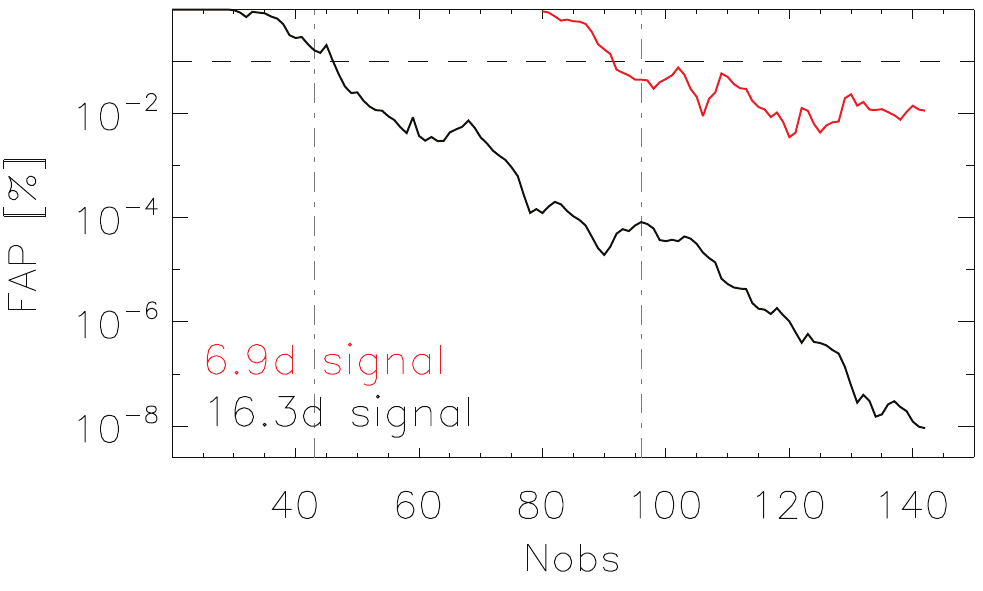}}
	\caption{FAP development for the 16.3 (black) and the 6.9\,day (red) signals of the RV data using an increasing number of data points. The dashed horizontal line shows the analytical 0.1\% FAP level, and the two vertical dash-dotted lines divide the data into the different seasons.}
	\label{FAP}
\end{figure}

In Fig.\,\ref{FAP} we show the time evolution of the FAPs of the 16.3- and 6.9-day signals in the RV data. We calculate the analytical FAP level of the most prominent peaks inside an interval of 16.3$\pm$1.5 and 6.9$\pm$1.0\,days, respectively, using an increasing number of data points. The 16.3-day signal shows a steady increase of significance over the entire observation time. The planetary signal reached a 1\% FAP level with approximately 60 observations, was hidden behind the noise until approximately 90 observations, and finally fell rapidly below a 0.1\% FAP level with 100 observations. The diagram shows changes in the significance of the signals between seasons 3 and 4, in agreement with the change of the activity pattern discussed above.

\subsection{Photometry} \index{de:pho} \label{de:pho}

As support to the HARPS-N spectroscopy, GJ\,3942 was monitored photometrically by the APACHE \citep[A PAthway toward the Characterization of Habitable Earths,][]{2013EPJWC..4703006S} and EXORAP (EXOplanetary systems Robotic APT2 Photometry) programs. In the following, we discuss the GLS analysis of this data together with additional photometry from WASP \citep[Wide Angle Search for Planets,][]{2010A&A...520L..10B} and Hipparcos \citep{1997A&A...323L..49P} catalogs, in order to identify at least the rotational periodicity at 16.3\,days.

\begin{figure}[tb]
	\resizebox{\hsize}{!}{\includegraphics[clip=true]{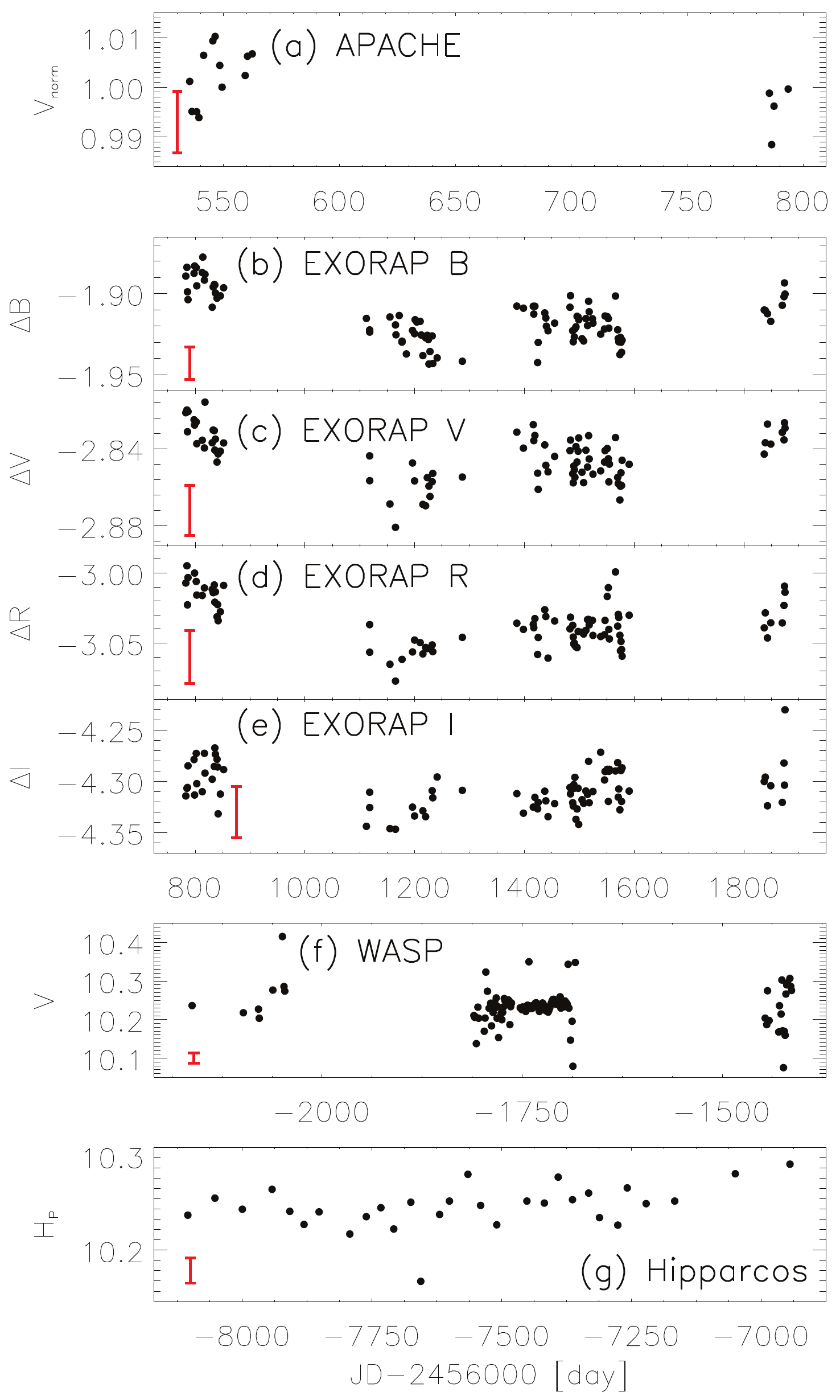}}
	\caption{Photometry of GJ\,3942 including normalized data from APACHE (panel a), magnitude differences (GJ\,3942 and calibration stars) of EXORAP filters $B$ (panel b), $V$ (panel c), $R$ (panel d), and $I$ (panel e), and visual magnitudes from WASP (panel f), and Hipparcos catalogs (panel g). The red lines on the lower left show as proxys for the uncertainties of the measurements the rms of the nightly averages (panels a, f, g), and the RV rms of the calibration stars (b to e). We note that APACHE was used during seasons 1 \& 2 of the HARPS-N observations, EXORAP from season 2 to 4 and beyond, and WASP and Hipparcos data are from around 2007 and 1991, respectively.}
	\label{Ph1}
\end{figure}

\begin{figure}[htb]
	\resizebox{\hsize}{!}{\includegraphics[clip=true]{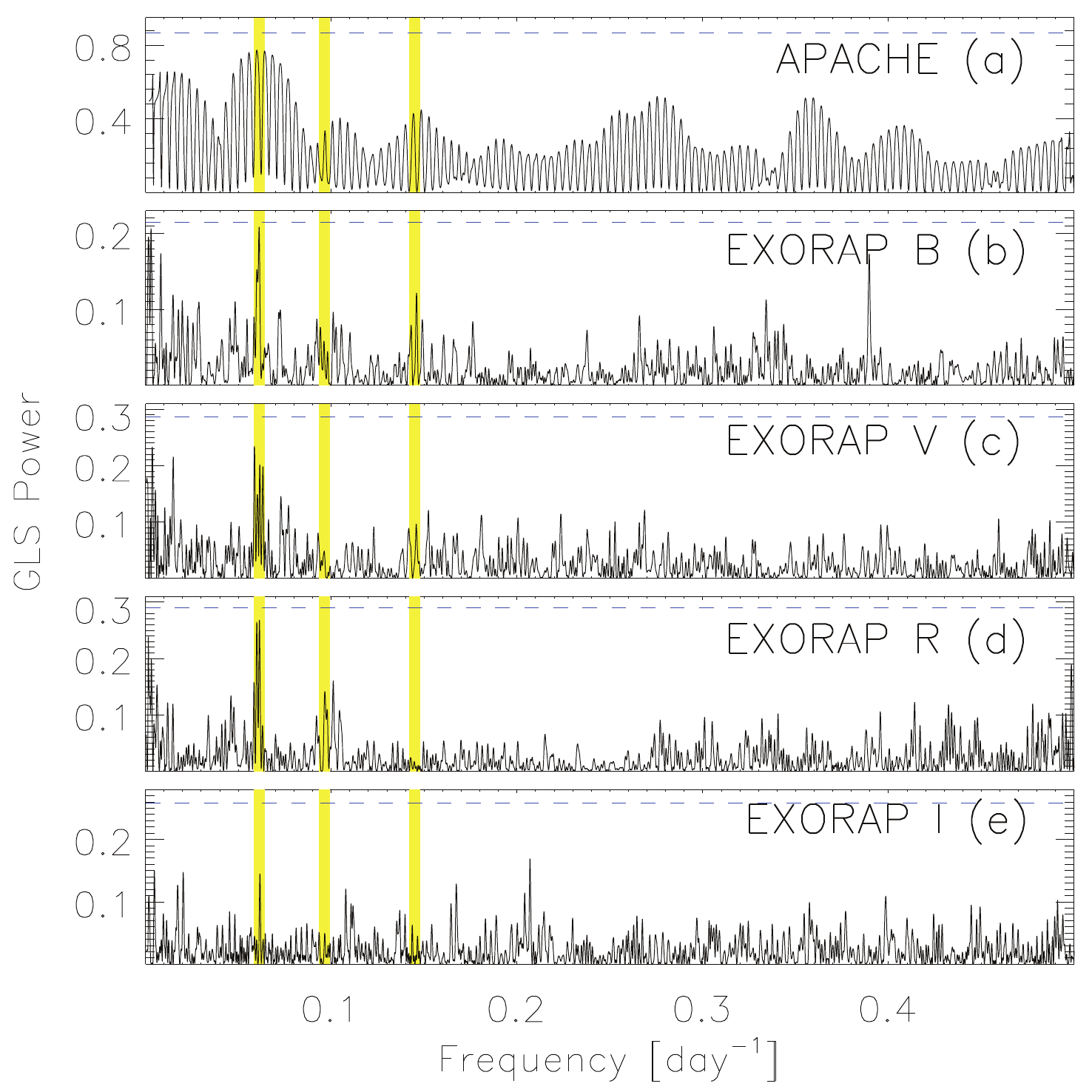}}
	\caption{GLS periodograms of the different photometric datasets including APACHE (panels a), and EXORAP filters $B$ (panel b), $V$ (panel c), $R$ (panel d), and $I$ (panel e). We mark by the yellow bands the periods at 16.3, 10.4, and 6.9\,days. The blue dashed horizontal line indicates the 1\,\% FAP level.}
	\label{Ph2}
\end{figure}

The APACHE photometric survey monitored GJ\,3942 with a 40~cm telescope located in the Astronomical Observatory of the Autonomous Region of the Aosta Valley (OAVdA, +45.7895 N, +7.478 E, 1650 m.a.s.l.), between August, 30$^{\rm th}$ and September, 29$^{\rm th}$ 2013 (12 nights of season 1 of HARPS-N observations), and May, 7$^{\rm th}$ to 15$^{\rm th}$ 2014 (4 nights of season 2). The dataset covers a timespan of 258\,days, and is composed of 418 useful measurements. The observations were carried out using a Johnson \textit{V} filter following the standard strategy used by APACHE, consisting in three consecutive exposures repeated at intervals of $\sim$20 to 25 minutes, while the target is $>$35$^{\circ}$ above the horizon. The images were reduced with the standard pipeline TEEPEE by the APACHE team \citep[see][]{2012MNRAS.424.3101G}. The light curve, including 141 data points was obtained by using an aperture of 6.5 pixels (9.75 arcsec) and a set of 5 comparison stars (UCAC4 716-054601, UCAC4 716-054610, UCAC4 715-054335, UCAC4 715-054346, and UCAC4 715-054359). To analyze the data with the GLS, we calculated nightly averages (see panel a of Fig.\,\ref{Ph1}) with a dispersion of 5.9\,mmag and average errors of 4.2\,mmag. The GLS periodogram of those 16 data points is shown in panel (a) of Fig.\,\ref{Ph2}. A best fit is found for a period of 18.1\,days and a semi-amplitude of 7.0\,mmag with a significance quite close to the 1\,\% threshold. 

EXORAP is carried out at INAF-Catania Astrophysical Observatory with a 80-cm f/8 Ritchey-Chretien robotic telescope (Automated Photoelectric Telescope, APT2) located at Serra la Nave (+14.973$^{\circ}$E, +37.692$^{\circ}$N, 1725 m a.s.l.) on Mt. Etna. $BVRI$ photometry of the star was collected over 85 nights between May, 5$^{\rm th}$ 2014 and April, 30$^{\rm th}$ 2017. The data cover season 2 to 4 of the HARPS-N observations. To obtain differential photometry, we started with an ensemble of $\sim$6 stars, the nearest to GJ\,3942 having similar brightness. Then we checked the variability of each of them by building their differential light curves using the rest of the sample as reference. That way we selected the 4 least variable stars of the sample. The rms of the ensemble stars is 10, 13, 19, 25~mmag in $B$, $V$, $R$, and $I$, respectively. We obtained 100/89/87/86 data points for $B/V/R/I$ filters (see panels b, c, d, and e of Fig.\,\ref{Ph1}), respectively, and no data points were rejected by a 5$\sigma$-clipping. We calculate average dispersions of 15.1/12.7/17.3/20.1\,mmag and average photometric uncertainties (sky + poisson) of 1.1/0.9/1.2/1.1\,mmag for the four filters. For every dataset, we correct for a long-period periodicity ($>$250\,days). The GLS periodograms of the residuals are shown in Fig.\,\ref{Ph2} for filters $B$ (panel b), $V$ (panel c), $R$ (panel d), and $I$ (panel e). The 16.3-day signal is seen in every dataset, but it is at the 1\% FAP level in the $B$ filter and very close to it in $V$ and $R$. Best fit periodicities are thereby 16.34/17.04/16.30\,days with semi-amplitudes of 6.5/6.6/7.8\,mmag in $BVR$ filters. Those values reproduce very well our previous estimations of the rotational period. The fits reduce the respective overall variation by 11.5/12.7/16.8\,\%. We do not show the WFs of all the photometric time-series data since they do not influence their respective periodograms in the important period range. 

We used WASP\footnote{\href{url}{http://exoplanetarchive.ipac.caltech.edu}} optical photometry containing 7\,447 data points. The target was observed in 120 nights from April, 12$^{\rm th}$ 2006 to April, 30$^{\rm th}$ 2008. We show the data after 1$\sigma$-clipping of the RVs and RV uncertainties and after averaging the values of each night in panel (f) of Fig.\,\ref{Ph1}. It shows a mean magnitude of $V$=10.234$\pm$0.044\,mag and a mean error of 13.2\,mmag. No significant periodicities are found in the GLS analysis. A last dataset of 16 nightly averages of 116 measurements from the Hipparcos catalog are shown in panel (g) of Fig.\,\ref{Ph1}. They were observed from January, 3$^{\rm rd}$ 1990 to March, 8$^{\rm th}$ 1993 in $H_{P}$ filter, which is similar to Johnson's $V$. We calculate an average magnitude of 10.242$\pm$0.021\,mag and an average uncertainty of 12.2\,mmag. The GLS analysis of the data does not show any significant signals.

The photometric data of the APACHE and EXORAP programs confirm the 16-day rotational period of GJ\,3942. All smaller signals shorter than $<$100\,days are related to this main periodicity and are corrected for after pre-whitening the rotational periodicity. For the EXORAP data, which cover same seasons as our spectroscopic HARPS-N observations, we can identify a minimum in brightness by season 3 (mid-2015 or JD$\sim$2457200\,days).

\section{Detailed analysis} \index{da} \label{da}

Our analysis of the available RV time-series data of GJ\,3942 suggests the presence of a planetary companion with an orbital period of 6.9 days and of another signal at a period of 16.3 days that we attribute to the stellar rotation period as traced by magnetic regions on the stellar surface. In the following, we compare different methods in order to find arguments for those findings and to further investigate whether the remaining signals could be suggestive of the presence of a second planet or rather are related to magnetic activity as arising from the evolution and decay of active regions and/or WF aliasing. Here we use the classical iterative method using a modified GLS code, the likelihood-ratio periodograms including a moving average term, and the Gaussian-process regression.

\begin{figure}[tb]
	\resizebox{\hsize}{!}{\includegraphics[clip=true]{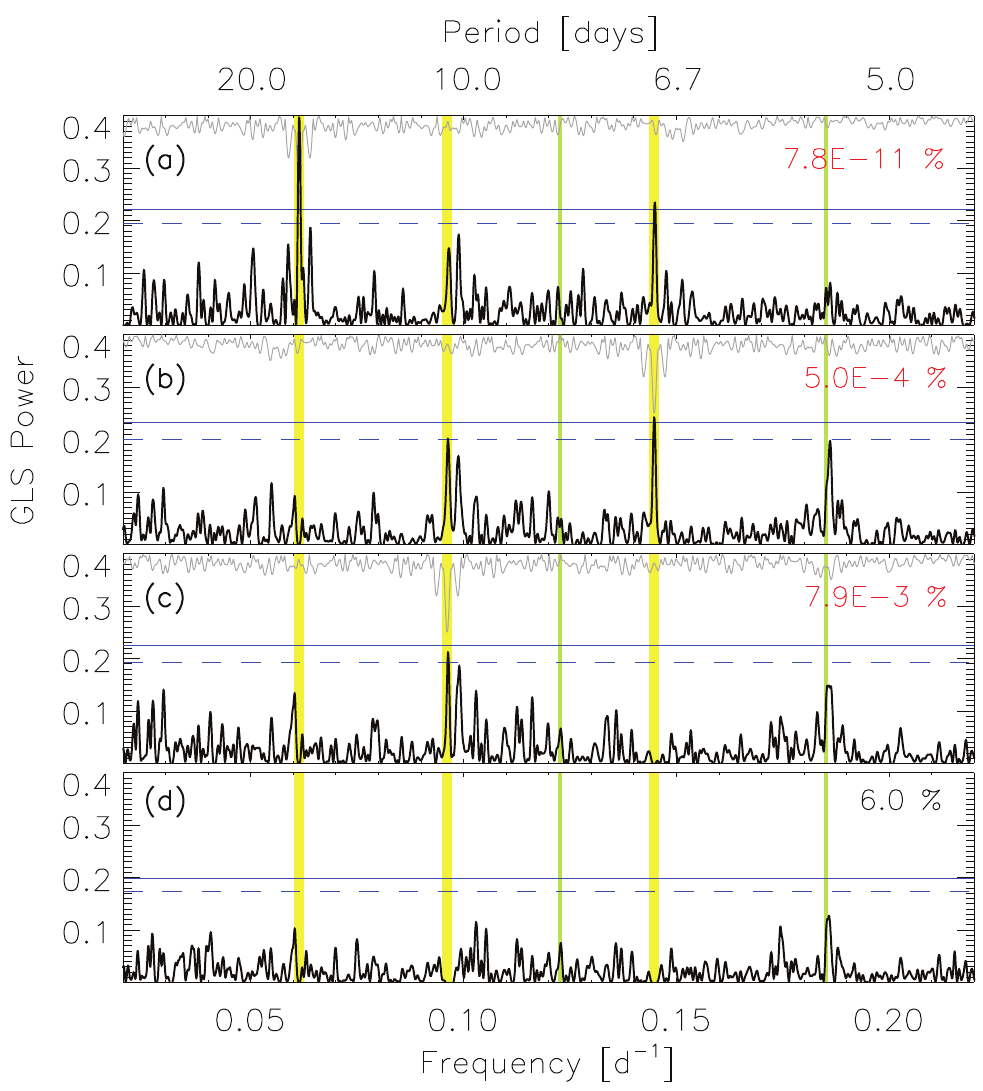}}
	\caption{GLS periodograms of the RV data of GJ\,3942. The contents of the panels are: (a) periodogram of the RV data; (b) residuals after correcting for a 16.3-day signal; (c) residuals after subtracting simultaneously 16.3- and 6.9-day signals; and (d) residuals after simultaneously correcting for signals at 16.3, 6.9, and 10.4\,days. For each panel, we give the analytical FAP of the respective model (in red if FAP$<$0.1\%). Yellow bands indicate the periods at 16.3, 10.4, and 6.9\,days. The green lines show the first and second harmonics of the 16.3-day signal, namely 8.2 and 5.4\,days. Blue horizontal lines show the 1\% (dashed) and 0.1\% (solid) FAP levels (from bootstrapping) of the respective dataset. Additionally, we show in the upper parts of each periodogram the WF in gray after being flipped, scaled, mirrored and period-shifted to the main peak of the respective dataset.}
	\label{c3e}
\end{figure}

\subsection{M-GLS analysis} \index{da:gls} \label{da:gls}

To further analyze the data, we eliminate the best sinusoidal fit to the most significant signal in the RV time series. We then recurrently continue this procedure with the residuals in an iterative way following a procedure called pre-whitening \citep[e.g.,][]{2013AN....334..616H}. Instead of correcting for each signal one after the other, we use a generalization of the GLS algorithm that considers a simultaneous multi-frequency fit, i.e., a GLS with more than one dimension, called ``Multi-dimensional GLS'' (M-GLS). In this way, the method employs an approach that mitigates cross-talk between signals having similar frequencies and amplitudes that could affect the classical pre-whitening technique.

\begin{figure}[tb]
	\resizebox{\hsize}{!}{\includegraphics[clip=true]{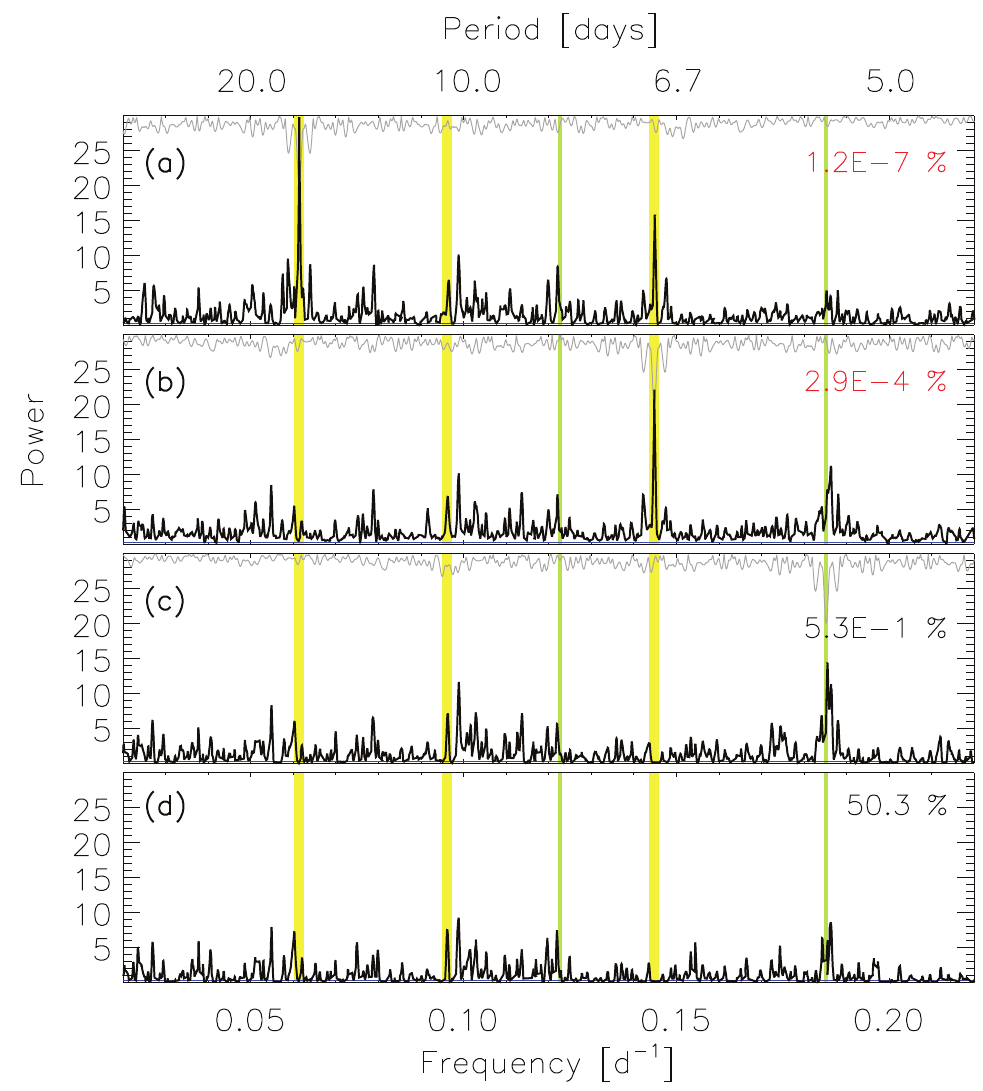}}
	\caption{Likelihood-ratio periodograms of the RV data of GJ\,3942. The contents of the panels are: (a) periodogram of the RV data; (b) residuals after correcting for a 16.3-day signal; (c) residuals after subtracting simultaneously 16.3- and 6.9-day signals; and (d) residuals after simultaneously correcting for signals at 16.3, 6.9, and 5.4\,days. The rest of the symbols are the same as in Fig. \ref{c3e}.}
	\label{c3f}
\end{figure}

We show in panel (a) of Fig.~\ref{c3e} the GLS periodogram of the RV data scaled to its main peak. Only the period range from 4.4 to 50 days (0.02 to 0.22\,days$^{-1}$) is plotted because of the relevance to this study and the lack of additional long-period signals. On top of the panel the WF is shown in gray. It is period-shifted and mirrored to the main peak of the periodogram, and flipped and scaled for display purposes. This way, it shows the influence of the combination of the time-sampling and the selected signal on the periodogram. Most importantly in our case it highlights yearly aliases seen on both sides of a peak. The sinusoidal fit parameters of the 16.3-day signal are shown in Table \ref{Res} under N$_{signal}$=1, and the GLS periodogram of the residuals, which have their rms reduced by 22\%, in panel (b). After pre-whitening the strongest signal, other features at short frequencies decrease in power but the 6.9-day peak does not vary. In addition, the two peaks at 10.1 and 10.4\,days increase their power level and reach a FAP of 1\%. There are also significant changes at around 5.4 days, and we hypothesize that this peak rising up to the 1\% FAP level can be related to the second harmonic of the 16.3-day signal, arising from the deviation of the rotational modulation effect from a purely sinusoidal shape. Panel (c) in Fig.~\ref{c3e} shows the periodogram after simultaneously correcting for the 16.3-day signal, with exact parameters shown in Table\,\ref{Res} under N$_{signal}$=2, and the next most prominent peak at 6.9\,days, which we attribute to GJ\,3942\,b. The rms of the RV data is reduced by a further 12\%. This procedure cleans the period at 6.9\,days efficiently but at the same time a yearly alias of the 16.3-day signal gains slightly in power. The 5.4-day signal becomes slightly less significant and the 10.4-day signal, now the most prominent peak, is close to reaching the 0.1\% FAP level. The same procedure is used to remove this third additional signal, reducing the RV rms by another 7\%. The fit uses the parameters shown in Table\,\ref{Res} with N$_{signal}$=3. The GLS periodogram of the residuals in panel (d) indicates that no statistically significant signal is left, although the strongest peak appears at 5.4\,days. We note that the subtraction of the 10.4-day signal also removes the one at 10.1\,days as they are yearly aliases of one another: (10.09$^{-1}$-10.38$^{-1}$)$^{-1} \approx$ 365\,days. The classical pre-whitening approach using the M-GLS algorithm therefore favors three periodic signals of 16.3, 6.9 and 10.4\,days.

\begin{table*}
	\caption{Best-fitting parameters of the RV data of GJ\,3942 and its residuals during the pre-whitening procedure of the three methods applied.}
	\label{Res}
	\begin{center}
		\begin{tabular}{lc|ll|lc}
			\hline \hline
			method & N$_{\rm signal}$ & P [day]             & K [\ms]             &  significance & rms [\ms]   \\ \hline
			\noalign{\smallskip}
			M-GLS & 1 &  16.29    & 5.1                    & FAP=7.8$\cdot$10$^{-11}$\%             & 4.65  \\
			& 2 & 16.28 / 6.91               & 4.7 / 3.3          & FAP=5.0$\cdot$10$^{-4}$\% & 4.08    \\
			& 3 & 16.27 / 6.91 / 10.38   & 4.7 / 3.0 / 2.6 & FAP=7.9$\cdot$10$^{-3}$\% & 3.79    \\ \hline
			MA   & 1 &  16.29                          & 5.6                & FAP=1.2$\cdot$10$^{-7}$\% & 3.98  \\
			& 2 & 16.28 / 6.91                 & 5.3 / 3.3        & FAP=2.9$\cdot$10$^{-4}$\% & 3.43    \\
			& 3 & 16.29 / 6.91 / 5.39     &  5.7 / 3.3 / 2.3 & FAP=5$\cdot$10$^{-1}$\%           & 3.14    \\ \hline
			GP   & 2 & (16.2 /) 6.91                            & (5.5 /) 3.1     & BIC=861.3  & 2.36    \\
			& 3 & (16.2 /) 6.91 / 20.44                & (5.8 /) 2.8 / 2.2        & BIC=856.7 & 1.70   \\
			\noalign{\smallskip}
			\hline \hline
		\end{tabular}
		\tablefoot{The parentheses in the first signal in the GP method indicate that the value is a result of the treatment of red noise. FAPs are analytical using Horne's formula.}
	\end{center}
\end{table*}

\subsection{Likelihood-ratio periodograms} \index{da:lik} \label{da:lik}

To further analyze the data independently, we use likelihood-ratio periodograms on our RV dataset. This includes considering Keplerian fits of planetary orbits and periodic stellar signals as well as a first order moving average (MA) component with exponential smoothing to account for the remaining correlated signals as red-noise \citep[see e.g.,][]{2014MNRAS.437.3540F, 2013A&A...549A..48T}. For the details of the likelihood-function which is calculated by the best fits to the time-series data, we refer to \cite{2016Natur.536..437A}. The additional term of the first order $MA$ accounts for the correlated noise considering the RV residual $\epsilon$ of the previous measurements at t$_{{\rm i}-1}$. The $MA$ at each time t$_{\rm i}$ is then given by
\begin{equation}
{MA}_{\rm i} = \phi \cdot \epsilon_{{\rm i}-1} \exp (\frac{{\rm t}_{{\rm i}-1  }-{\rm t}_{\rm i}}{\tau}).
\end{equation}
The coefficient $\phi$ measures the correlation of RVs at t$_{\rm i}$ and t$_{{\rm i}-1}$ and $\tau$ is a characteristic time decay measuring the impact of that correlation to the RVs over time.

In Fig.~\ref{c3f} we show equivalent panels to those in Fig.~\ref{c3e} using the likelihood-ratio periodograms. The plots are again scaled to the main peak of the periodogram in panel (a). In general, thanks to the more sophisticated treatment of data correlations, the background noise level is diminished when using this technique. We show the different parameters used for the fits in Table\,\ref{Res} and the additional parameters used by this technique in Table\,\ref{Res2}. Panel (a) shows two strong peaks, as before, at approximately 16.3 and 6.9\,days, weaker double peaks at around 10\,days, and a moderately strong periodic signal at 8.2\,days, which is interpreted as the fist harmonic of the 16.3-day period. As can be seen in panel (b), removing the 16.3-day signal diminishes the significance of its first harmonic and increases the power of the 6.9-day signal, as do again signals at approximately 5\,days. The RV rms is reduced in this step by 33\%. Removing both the 16.3- and 6.9-day signals reduces the rms by another 14\% and the 5.4-day peak, which has some structure, becomes the most significant one, as shown in panel (c). Subtracting the 5.4-day signal leaves still some power at a similar period of 5.2\,days, as can be seen in panel (d). Also, we have some excess power left at approximately 8, 10, 16 and 20\,days. The reduction of the rms by this last step is of the order of 8\%.

The MA method seems to favour a one-planet model with a slightly lower FAP when compared to the M-GLS method and small orbital and effective eccentricities, respectively. Together with the red-noise treatment, the method is thereby able to reduce the rms of the residuals significantly in comparison with the M-GLS analysis. The signal at 5.4\,days appears strong after correcting for the rotational periodicity in (b) indicating again that it might be the second harmonic of the 16.3-day signal. A further remaining peak at 5.2\,days in (d) could then represent a first harmonic of the signal at 10.4\,days that we find with the M-GLS method.

\begin{table}
	\caption{Additional best-fitting parameters of the RV data of GJ\,3942 and its residuals using the MA method.}
	\label{Res2}
	\begin{center}
		\begin{tabular}{l|lll}
			\hline \hline
			N$_{\rm signal}$ & $\phi$ [\ms] & $\tau$ [d] & eccentricity \\ \hline
			\noalign{\smallskip}
			1 & 1.10 & 0.53 & 0.23  \\
			2 & 1.33 & 0.67 & 0.20 / 0.12  \\
			3 & 0.92 & 0.26 & 0.20 / 0.03 / 0.27 \\
			\noalign{\smallskip}
			\hline \hline
		\end{tabular}
	\end{center}
\end{table}

\subsection{Gaussian process regression} \index{da:gau} \label{da:gau}

\begin{figure*}%[!htbp]
	\resizebox{\hsize}{!}{\includegraphics[width=9.5cm]{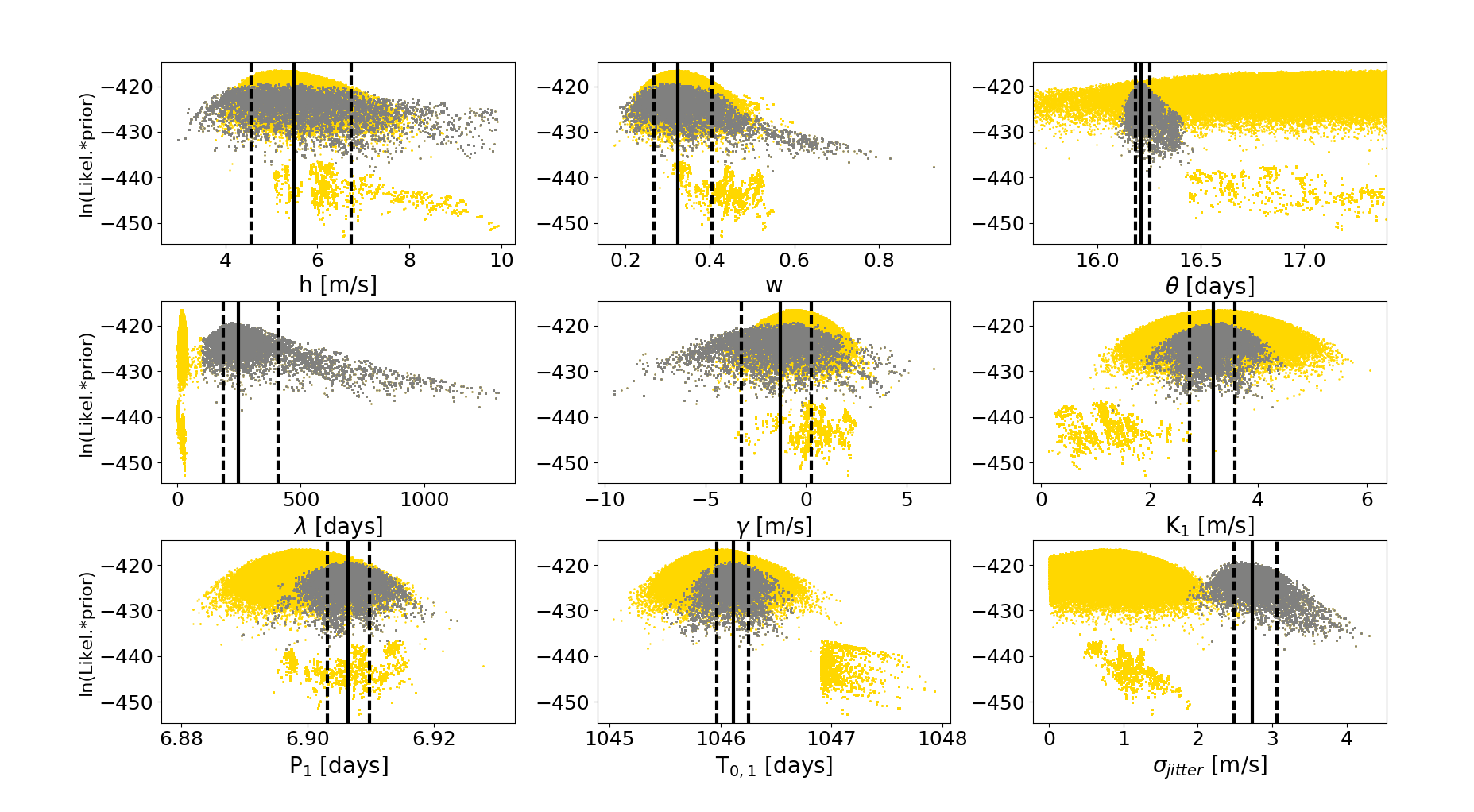}}
	\caption{Distributions of the total MCMC samples for the jump parameters of the GP one-planet model as function of the natural logarithm of the likelihood function (yellow dots). The samples corresponding to $\lambda$>100 days are shown as gray dots. The vertical lines represent the median values (solid) and the 16th and 84th percentiles (dashed) of the latter distribution, which are listed in Table\,\ref{priors1plan}. }
	\centering
	\label{lnLikevsparam1planet}%
\end{figure*}

Following our multi-technique approach, we modeled the RVs of GJ\,3942 by using Gaussian process (GP) regression through a Markov chain Monte Carlo (MCMC) algorithm. This method is becoming increasingly popular as an efficient and physically robust way to mitigate signals in RV data that can be ascribed to stellar activity, while retrieving reliable estimates of the planetary orbital parameters. For details of how GPs have been used for the analysis of RV datasets, we refer to, e.g., \cite{2014MNRAS.443.2517H, 2015ApJ...808..127G, 2015MNRAS.452.2269R, 2016A&A...593A.117A, 2016A&A...588A..31F, 2016AJ....152..204L, 2017A&A...598A.133D, 2017A&A...599A.126D, 2017AJ....153....9C, 2017arXiv170605459A}. In this work we have adopted the widely-used quasi-periodic kernel described by the following covariance matrix:
\begin{eqnarray} \label{eq:eqgpkernel}
	k(t, t^{\prime}) = h^2\cdot\exp\bigg[-\frac{(t-t^{\prime})^2}{2\lambda^2} - \frac{sin^{2}(\frac{\pi(t-t^{\prime})}{\theta})}{2w^2}\bigg] + \nonumber \\
	+\, (\sigma^{2}_{\rm instr, RV}(t)\,+\,\sigma_{\rm inst,jit}^{2})\cdot\delta_{\rm t, t^{\prime}},
\end{eqnarray}
where $t$ and $t^{\prime}$ indicate two different epochs. This functional form is particularly suitable to model periodic short-term stellar signals, modulated by the stellar rotation period, by allowing for an exponential decay of the correlations between different epochs. The hyper-parameters appearing in Eq.\,\ref{eq:eqgpkernel} are linked to some of the physical phenomena underlying the stellar noise: $h$ represents the amplitude of the correlations, $\theta$ generally represents the rotation period of the star, $w$ is the length scale of the periodic component, linked to the size evolution of the active regions, and $\lambda$ is the correlation decay timescale that we assume to be related to the active regions lifetime. Moreover, $\sigma_{\rm RV}(t)$ is the RV internal error at time \textit{t}, $\sigma_{\rm jit}$ is the additional uncorrelated ``jitter'' term that we add in quadrature to the internal errors to take into account instrumental effects and other noise sources included neither in $\sigma_{\rm RV}(t)$ nor in our stellar activity framework, and $\delta_{\rm t, t^{\prime}}$ is the Dirac delta function.

We refer to \cite{2017A&A...599A.126D} for the log-likelihood function to be maximized by the MCMC procedure and the general form for the models that we tested in this work. The selection of the model, multiplicity and the priors of the GP hyper-parameters are based on the outcome of the M-GLS and MA analyses of the previous sections.

\subsubsection{One-planet model} \index{da:gau:1D} \label{da:gau:1D}

We list in Table\,\ref{priors1plan} the uniform prior probability distributions for the parameters of the one-planet model. Regarding the GP hyper-parameters, we constrained \textit{w} between 0 and 1 from the discussion in \cite{2016AJ....152..204L}, $\theta$ was sampled in a range around the stellar rotation period of 16.3 days, and for $\lambda$ we adopted an upper limit with a value slightly larger than the dataset timespan. The planetary orbital period $P_{\rm 1}$ was sampled in a small interval around 6.9 days. The parameter space was explored using 150 random walkers. We randomly initialized the positions for the $\lambda$ GP hyper-parameter around 100 days using a Gaussian distribution ($\sigma$=50 days) to start the exploration of the parameter space within a sufficiently wide range for the correlation decay timescale. The progress of the MCMC fitting procedure was monitored by evaluating the Gelman-Rubin convergence parameters as defined by \cite{2006ApJ...642..505F}. The final distributions of the free parameter samples as a function of $\ln$(likelihood*prior) are shown in Fig.\,\ref{lnLikevsparam1planet} by the yellow dots These were obtained by applying a first burn-in of 3\,000 steps and then discarding additional samples up to the first MCMC step at which all chains have had at least one value of $\ln$(likelihood*prior) lower than the median of the $\ln$(likelihood*prior) dataset, following the prescription of \citet[][and references therein]{2013PASP..125...83E}. We note that the posterior distribution of the $\lambda$ timescale appears to be bi-modal. The maximum a posteriori probability estimate $\lambda$=17.1 days, which is the expected peak at about the stellar rotation period. Of astrophysical relevance is the second local maximum at $\sim$250 days \citep{2009A&ARv..17..251S,2015PhDT.......177D}. For deriving the final best-fit parameter values, we then consider only samples for which $\lambda$>100 days (the period where both solutions meet), resulting into the posterior distributions shown in Fig.\,\ref{lnLikevsparam1planet} by the gray dots. The uncorrelated jitter appears to be bimodal whereas for our 250\,day $\lambda$ the solution occurs at $\sim$2.7\,\ms, which is more than double the mean RV error. The high value for the additional jitter could be indicative of the existence of a second planetary signal in the dataset. Table\,\ref{priors1plan} shows the corresponding best-fitting values and uncertainties for each jump parameter, calculated as the median of the marginal posterior distributions and the 16\% and 84\% quantiles. In Table\,\ref{Res} we show the characteristics of the fit to compare them to the other methods. We note that the $\theta$ parameter from the GP term is regarded as a period of the first signal, but given in parentheses. 

\begin{figure}%[!htbp]
	\resizebox{\hsize}{!}{\includegraphics[width=9.5cm]{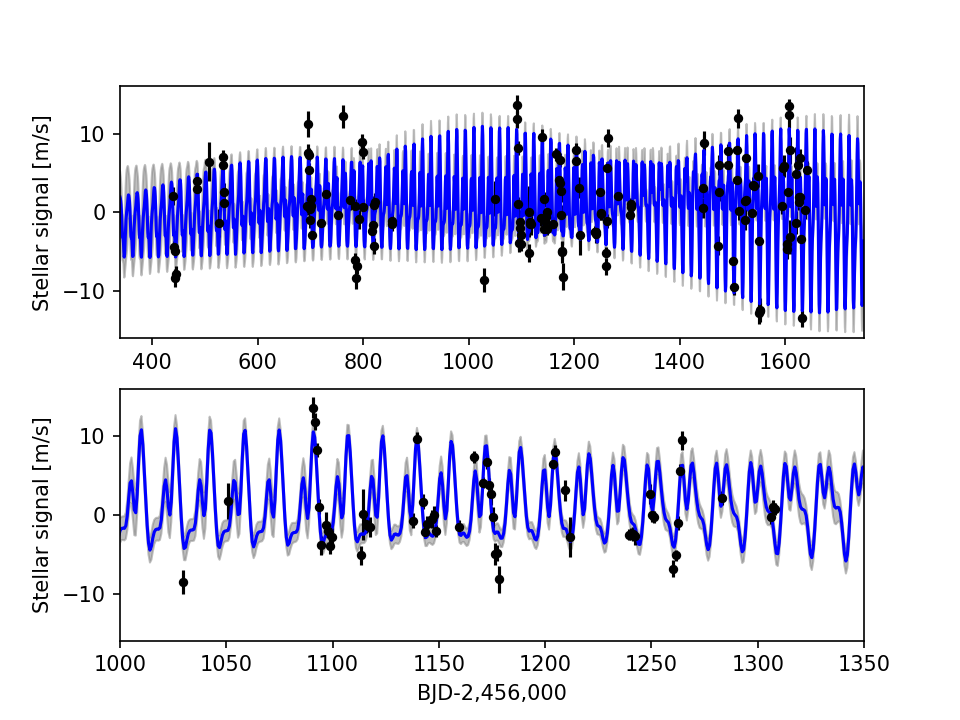}}
	\caption{Radial velocity residuals time series (black dots), after subtracting our best-fit orbital solution for GJ\,3942\,b. The blue line with gray shaded 1$\sigma$ regions represents our best-fit GP quasi-periodic model for the correlated stellar noise. The top plot shows the complete dataset, while the lower plot shows a blow-up of the third epoch for easier visualization of the agreement between the model and the data.}	
	\centering
	\label{noisemodel_lambda_full}
\end{figure}

The values of the one-planet model reproduce the results of the other methods very well. The value for $\lambda$ confirms the described changes of magnetic activity phenomena on a seasonal time-scale. Figure\,\ref{noisemodel_lambda_full} shows the stellar contribution to the RV after removing the best-fit planetary solution. We note a slight increase of the scatter over the four seasons, corresponding to the RV values. The residuals of our global model show an rms of 2.36\,\ms. The Bayesian strength of the models tested in this work are evaluated by using the Bayesian Information Criteria (BIC), and the results are interpreted by adopting the empirical scale presented in \cite{raftery95}. For the one-planet model we obtain BIC=861.3.

\begin{table}
	\caption[]{Uniform prior probability distributions and best-fitting estimates for the hyper-parameters used in the one-planet circular model. The adopted best-fit values were calculated from samples for which $\lambda$>100 days (see text for explanation).}
	\label{priors1plan}
	\small
	\centering
	\begin{tabular}{l|lll}
		\hline \hline
		Jump parameter     & > bound & $<$ bound  & Best fit\\ \hline
		\noalign{\smallskip}
		h [m$\,s^{-1}$] & 0.1 & 10 & 5.5$^{+1.2}_{-0.9}$ \\
		$\lambda$ [days] & 0.1 & 1300  & 249$^{+160}_{-62}$  \\
		$w$ & 0 & 1 &  0.32$^{+0.08}_{-0.06}$ \\
		$\theta$ [days] & 15 & 19 & 16.21$^{+0.04}_{-0.03}$ \\ \hline
		$\gamma$ [m$\,s^{-1}$] & -10 & +10 & -1.3$^{+1.5}_{-1.9}$\\
		$\sigma_{\rm jitter}$ [m$\,s^{-1}$] & 0 & 10 & 2.7$^{+0.3}_{-0.2}$\\
		$K_{\rm 1}$ [m$\,s^{-1}$] & 0.1 & 10 & 3.2$\pm$0.4  \\
		$P_{\rm1}$ [days] & 6 & 8 & 6.906$\pm$0.003 \\
		$T_{\rm 1,0}$ [JD-2\,456\,000] & 40 & 48 & 46.11$^{+0.13}_{-0.15}$\\
		\noalign{\smallskip}
		\hline \hline
	\end{tabular}
\end{table}

\subsubsection{Modeling activity index time series} \index{da:gau:mo} \label{da:gau:mo}

\begin{figure}%[!htbp]
	\resizebox{\hsize}{!}{\includegraphics[width=9.5cm]{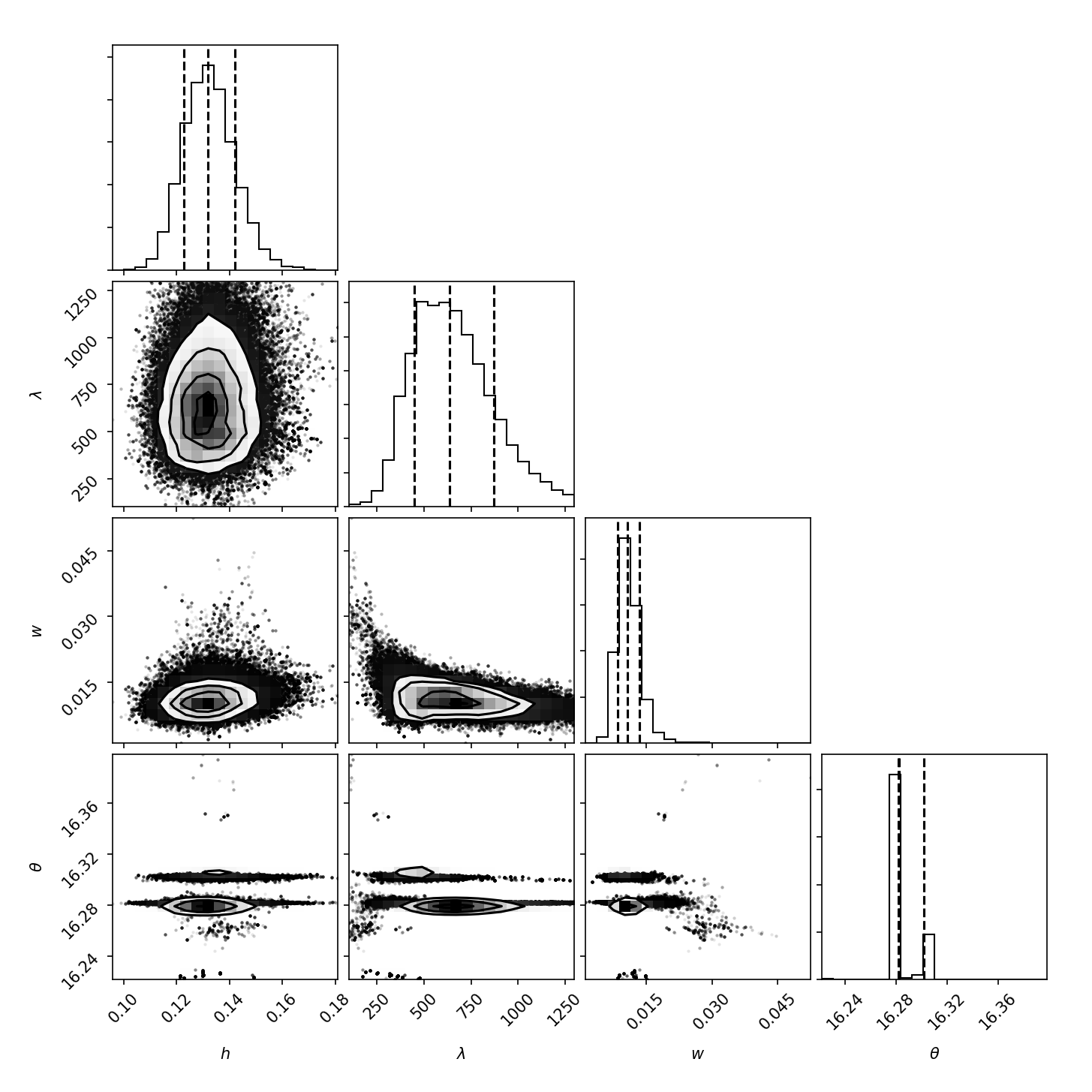}}
	\caption{Posterior distributions of the GP hyper-parameters for the quasi-periodic kernel applied to the $S$ index time series. Dashed vertical lines mark the median values and the 16th and 84th percentiles of these distributions, which are listed in Table\,\ref{priorspostdistcahk}.}
	\centering
	\label{cahkpostdist}%
\end{figure}

The same approach was used to analyze the time series of the $S$ and H$\alpha$ indices. We performed a GP regression without including an additional jitter term in the kernel, and loosely conditioning the model \textit{a priori}. The hyper-parameter $\theta$ was constrained within a small range around a 16.3-day period, while we left $\lambda$ free. The positions of the random walkers were initialized at around 550 days. The uniform priors are listed in Table\,\ref{priorspostdistcahk}. The 50 independent chains reached convergence according to the criteria defined by \cite{2006ApJ...642..505F}. We show in Fig.\,\ref{cahkpostdist} the posterior distributions obtained after a burn-in as explained in the previous section. We see that the stellar rotation period $\theta$ appears to be slightly bi-modal but still tightly constrained around 16.3 days, thus reproducing the solution found fitting a simple sinusoid. The best-fit solution has very similar properties of that found for the stellar contribution in the RVs alone but with a larger $\lambda$ of 634\,days. Table\,\ref{priorspostdistcahk} lists the corresponding best-fitting values and uncertainties for each jump parameter.

\begin{table}
	\caption[]{Uniform prior probability distributions and best-fitting estimates for the hyper-parameters of the quasi-periodic kernel used to model the $S$ index dataset.}
	\label{priorspostdistcahk}
	\small
	\centering
	\begin{tabular}{l|lll}
		\hline \hline
		Jump parameter     & $>$ bound & $<$ bound & Best fit \\ 	\hline
		\noalign{\smallskip}
		h [\ms] & 0 & 0.5 & 0.13$^{+0.010}_{-0.009}$ \\
		$\lambda$ [days] & 0 & 1300 & 634$^{+236}_{-188}$ \\
		$w$ & 0 & 1 & 0.011$^{+0.003}_{-0.002}$ \\
		$\theta$ [days] & 15.5 & 19 & 16.2821$^{+0.0198}_{-0.0005}$\\
		\noalign{\smallskip}
		\hline \hline
	\end{tabular}
\end{table}

Applying the quasi-periodic kernel to the H$\alpha$ index dataset does not produce the same results as for the $S$ index. By adopting the same priors, the chains did not reach convergence, with $\theta$=16.262$\pm$0.007 days, and $\lambda$ peaking at the upper edge of the prior interval. This indicates very stable fit parameters over the time of observations. The result is in agreement with the lowest FAP of this index in Fig.\,\ref{F8b} {for the rotational signal, which indicates that it is best described by a simple sinusoid. It also agrees with the smallest variation of the fit parameters of Season 3 in Fig.\,\ref{F9}.

\subsubsection{Two-planet model} \index{da:gau:2D} \label{da:gau:2D}

As discussed before, the analysis of the RV residuals suggests the existence of an additional signal. To explore the significance of this possibility, we tested a model with two circular Keplerian signals by using the same uniform priors of the one-planet model, and a third periodicity between 0.1 and 25\,days (see Table\,\ref{priors2plan}). This time we used 100 independent MCMC chains, initially spread over a large fraction of the parameter space, and we stopped the run after 200,000 steps, without reaching the formal convergence. The results show that the posterior distribution for $P_{\rm2}$ is characterized by two local maxima at approximately 10.1 and 20.4 days, with the latter period being more significant. The solution points towards a similar evolutionary timescale $\lambda$ as for the one-planet model, and a slightly smaller semi-amplitude for the inner planet. The residuals have an rms of 1.70\,\ms, well below the expected jitter of 2.3\,\ms. In the residuals the 10.1-day periodicity is still visible. The BIC value for this model is 856.7. By using the approximation $BIC_{2}-BIC_{1} \simeq 2 \ln B_{12}$, we obtain $B_{12} \sim 0.1$ for the Bayes factor, indicative of a mildly positive evidence for the two-planet model. Models with free eccentricity did not converge.

\begin{table}
	\caption[]{Uniform prior probability distributions and best-fitting estimates for the hyper-parameters used in the two-planet circular model.}
	\label{priors2plan}
	\small
	\centering
	\begin{tabular}{l|lll}
		\hline \hline
		Jump parameter     & $>$ bound & $<$ bound & Best fit \\ \hline
		\noalign{\smallskip}
		h [m$\,s^{-1}$] & 0.1 & 10  & 5.8$^{+1.3}_{-1.0}$ \\
		$\lambda$ [days] & 0.1 & 1300  & 185$^{+63}_{-42}$ \\
		$w$ & 0 & 1 & 0.33$^{+0.06}_{-0.05}$\\
		$\theta$ [days] & 15 & 19 & 16.23$^{+0.04}_{-0.03}$ \\ \hline
		$\gamma$ [m$\,s^{-1}$] & -10 & +10 & -0.8$^{+1.6}_{-2.0}$ \\
		$\sigma_{\rm jitter}$ [m$\,s^{-1}$] & 0 & 10 & 2.1$^{+0.5}_{-0.3}$\\
		\noalign{\smallskip}
		$K_{\rm 1}$ [m$\,s^{-1}$]  & 0.1 & 10 & 2.8$\pm$0.4 \\
		$P_{\rm1}$ [days]  & 6 & 8 & 6.907$\pm$0.004\\
	$T_{\rm 1,0}$ [JD-2\,456\,000] & 40 & 48 & 46.12$\pm$0.16\\
	$K_{\rm 2}$ [m$\,s^{-1}$] & 0.1 & 10 & 2.2$\pm$0.4\\
	$P_{\rm2}$ [days] & 0.1 & 25 & 20.44$^{+0.05}_{-10.33}$ \\
	$T_{\rm2,0}$ [JD-2\,456\,000]  & 40 & 67 & 49.07$^{+0.77}_{-3.47}$ \\
	\noalign{\smallskip}
	\hline \hline
\end{tabular}
\end{table}

\section{Discussion} \index{di} \label{di}

Two main periodic signals in our data are confirmed by all the methods that we employ, one that is related to stellar magnetic activity and one that is best described as induced by the Keplerian motion of an orbiting planet. The differences between the various analysis methods appear in the strength and interpretation of subsequent signals, most precisely of a third RV periodicity that could potentially be an additional planet in the system (see Table\,\ref{Res}). 

The M-GLS fit proposes such second planetary candidate with an orbital period of 10.4\,days. But we model rotational and planetary modulations as simple sinusoids with constant parameters, although they might have evolving characteristics and eccentric orbits, respectively. The method also does not treat the correlated or uncorrelated noise components which make this simple correction, using the pre-whitening approach, more unreliable. The likelihood-ratio MA analysis instead includes a noise contribution into the fits. It favors an interpretation of harmonics of the main signal at 16.3\,days (P) instead of the existence of an additional independent signal. This is seen in panels (a) of Fig.\,\ref{c3e} for the 6.9-day and P/2 (8.15\,days) signals and in panel (c) for the 10.4-day and P/3 (5.4\,days) signals. Thus, the non-significant 5.4-day peak selected as third periodicity by this method would be a harmonic of the rotation period and the presence of such harmonics be a consequence of the varying and non-sinusoidal shape of the 16.3-day modulation in the RV time-series data. Also, we suspect some correlation between the time sampling and the periodic signals found, since we can reproduce the 5.4\,day peak by simple models simulating a dataset with mildly-eccentric 16.3, 10.4, and 6.9-day Keplerian signals. The GP regression method adjusts the observed data to a certain model of the stellar system and results in the detection of a 20.4-day period in addition to the strong 16.3 and 6.9-day signals. The suggestive harmonic chain at 20.4, 10.2, 6.8, and 5.1\,days, which are close to some of the significant signals that we have found throughout our study, deserves further analysis.

In the case of an active star, the presence of differential rotation, i.e., latitudinal differences in the rotation velocity, can manifest itself through the appearance of additional signals in the periodogram. The analysis of Kepler data \citep{2016MNRAS.463.1740B} shows that differential rotation is common in all stars but especially so in slow and moderate rotators, such as GJ\,3942. For example, \cite{2015A&A...583A..65R} used the $\alpha$ parameter ($\alpha=1-P_{\rm min}/P_{\rm max}$) to quantify the level of differential rotation through the empirical analysis of 12\,000 stars with photometry from the Kepler mission. We could hypothesize that the 20.4-day signal revealed by the GP method arises from active regions at high-latitude locations. Indeed, the $\alpha$ parameter of this period compared with 16.3\,days is 0.201, in agreement with Fig.\,9 of \cite{2015A&A...583A..65R}. There is, however, a strong indication that the 20.4-day signal and its harmonics are not associated to stellar activity. The reason is the clean periodogram of the activity indicators $S$ and $H\alpha$, where only the 16.3-day signal and, possibly, its much weaker first and second harmonic signals are seen. There are no traces of signals at 20.4\,days or any of its harmonics. Of course, using this argument to rule out the 20.4-day harmonic chain as arising from stellar activity hinges on the fact that the chromospheric emissions traced by these indicators are good proxies for photospheric active regions that produce the RV shifts. This seems, however, a reasonable assumption as such connection is observed in the Sun and we remind that we also do not see any of the periodicities in the photometric datasets but the 16.3-day periodicity.

Another scenario could be that the harmonic chain is produced by a very eccentric planet. For illustration, we show in the top panel of Fig.\,\ref{k2} the RV residuals of our MA method after correcting for the periodic signal at 16.3\,days and phase folded at a period of 20.52\,days, which is the best fit period in an interval from 20 to 21\,days. The resulting sinusoidal modulation has a semi-amplitude of only 1.88\,\ms, which is well below the expected uncorrelated RV jitter of 2.3\,\ms, and does not show any signs for high eccentricity. We further carried out a simulation analysis using the observed time-series dates and injecting signals at 16.3 and 20.4\,days and considering Gaussian errors. After performing 10\,000 trials by varying eccentricities for the rotational and planetary signals from 0 to 0.3 and from 0.3 to 0.9, we did not find any configuration that could reasonably reproduce the set of signal harmonics. From this, we deem the 20.4-day period and its harmonic chain as quite unlikely to arise from a highly-eccentric planet, or actually by more than one, since at those close periods the system would be dynamically unstable.

\begin{figure}[tb]
	\resizebox{\hsize}{!}{.\includegraphics[clip=true]{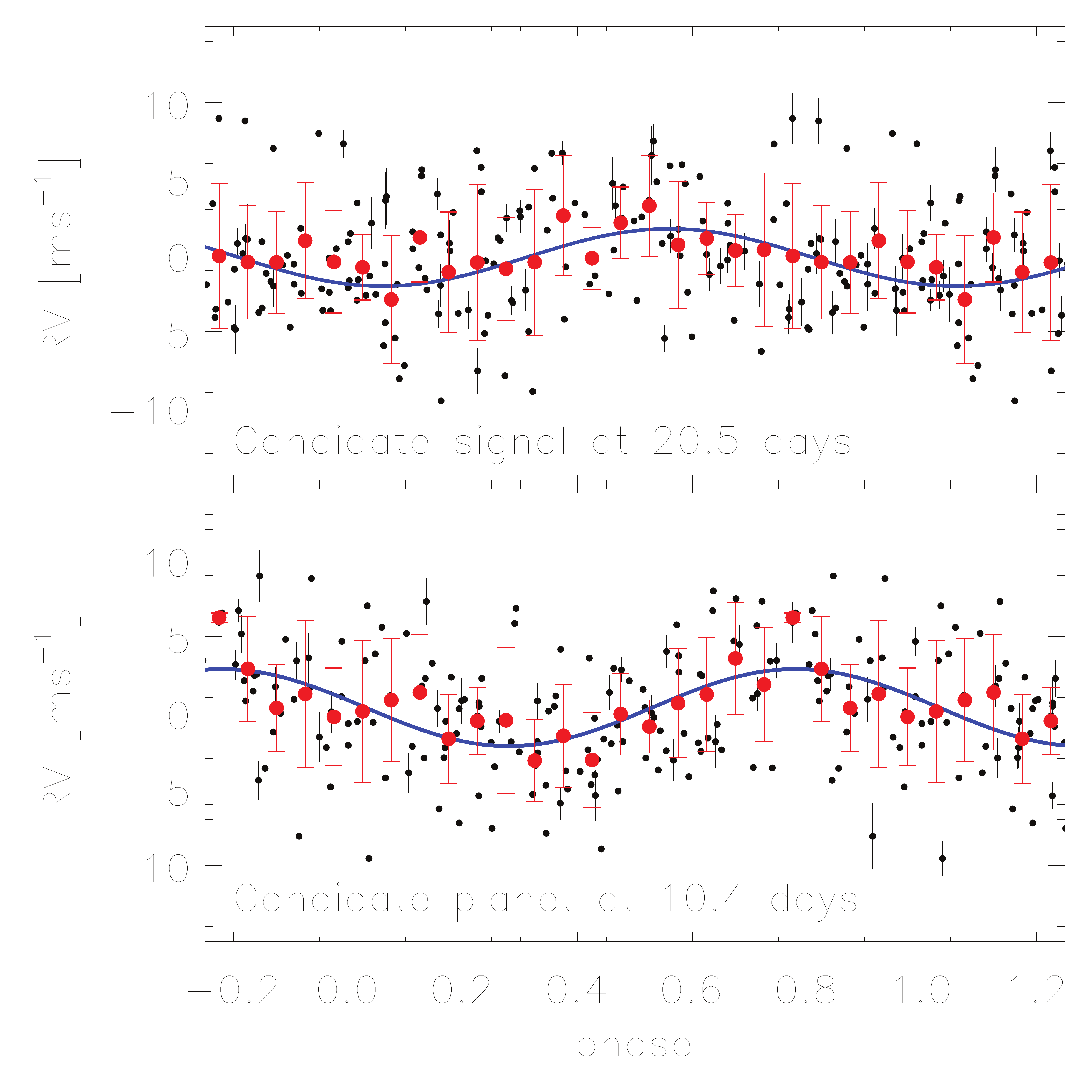}}
	\caption{Residuals and errors of the RVs in black dots folded to a period of 20.5 (top panel) and 10.4\,days (bottom panel) with semi-amplitudes of 1.88 and 2.52\,\ms, respectively. We used the data after pre-whitening the 16.3-day signal (top) and the 16.3 and 6.9-day signal (bottom) using the MA method.  The red dots correspond to averages in 0.05 phase intervals. We show the best sinusoidal fits in blue.}
	\label{k2}
\end{figure}

\begin{figure}[tb]
	\resizebox{\hsize}{!}{.\includegraphics[clip=true]{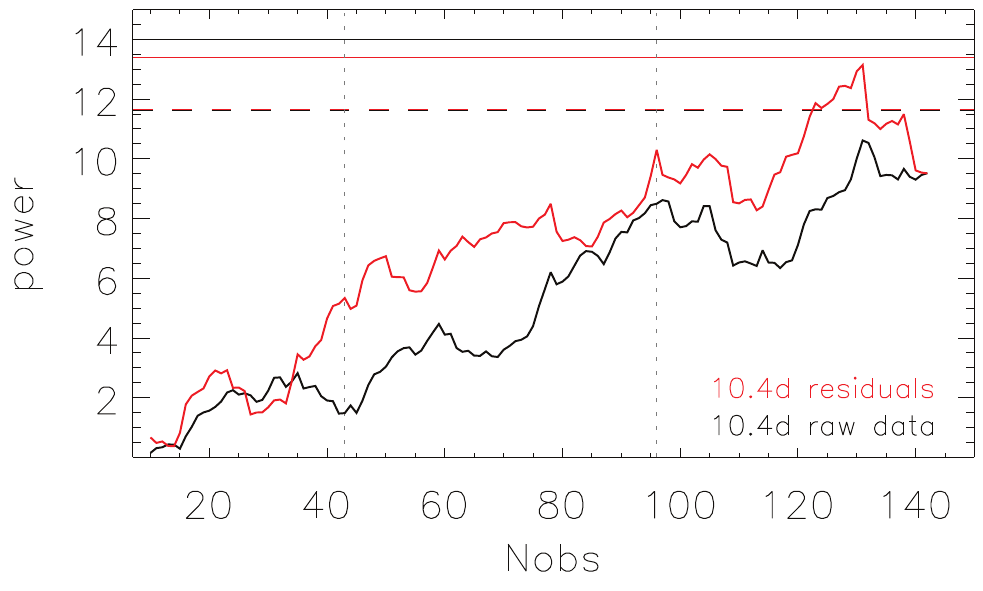}}
	\caption{Evolution of periodogram power of the signal at around 10.4\,days with the number of observations. We use the residuals after the subtraction of the 16.3- and 6.9-day signals of the best fit following our MA method in red and in the original data in black. The solid and dashed lines correspond to the 0.1 and 1\,\% FAP levels, respectively.}
	\label{f4ii}
\end{figure}

\begin{figure}[tb]
	\resizebox{\hsize}{!}{.\includegraphics[clip=true]{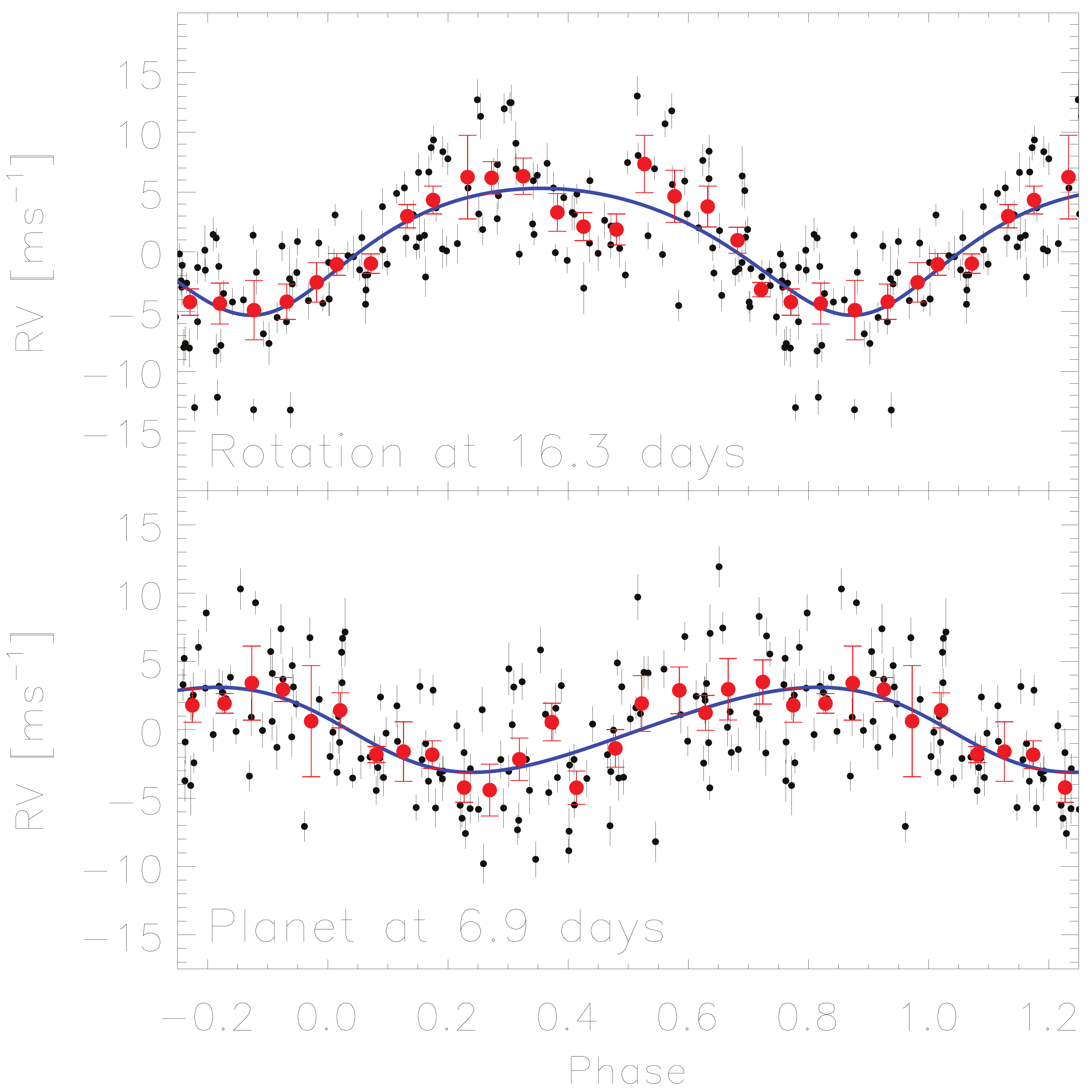}}
	\caption{Best simultaneous Keplerian fits (blue curve) following our MA method for periods of 16.28 (top) and 6.91\,days (bottom). The fits have semi-amplitudes of 5.3 and 3.3\,\ms, and eccentricities of 0.20 and 0.12, respectively. The data are shown without the contribution from the MA term.}
	\label{k1}
\end{figure}

\begin{table*}
	\caption{Summary of the characteristics of the RV periodicities of GJ\,3942 following the MA method.}
	\label{Tabresults}
	\begin{center}
		\begin{tabular}{ll|ll|l}
			\hline \hline
			\noalign{\smallskip}
			 & RVFIT & GJ\,3942\,b   & GJ\,3942\,c      & rotational\\
			 &    bounds    &            &     candidate	&   signal \\ \hline
			\noalign{\smallskip}
			P [days]   & fixed  & 6.905$\pm$0.040 & 10.378$\pm$0.089 & 16.283$\pm$0.220  \\
			K [\ms]    & 0 -- 15  & 3.29$\pm$0.12 & 2.52$\pm$0.07 & 5.31$\pm$0.07   \\
			$\omega$ [deg] & 0 -- 360 & 112.7$\pm$0.1 & -- & 188.0$\pm$1.4   \\
			e & 0 -- 0.75 & 0.121$\pm$0.118 & -- & 0.202$\pm$0.010    \\ \hline
			$M_{\rm p}\sin i$ [M$_{\oplus}$] & & 7.14$\pm$0.59 & (6.33$\pm$0.50)  & --  \\
			a [AU]  &        &  0.0608$\pm$0.0068 &  (0.0798$\pm$0.0089)  & -- \\
			\noalign{\smallskip}
			\hline \hline
		\end{tabular}
	\end{center}
\end{table*}

A further interesting hypothesis is that the signals are induced by a true planetary system of two or more planets in near resonant orbits. Orbital resonances are found to be quite common in exoplanet systems, including the TRAPPIST-1 system, with up to seven planets near mean-motion resonances \citep{2017Natur.542..456G, 2017arXiv170304166L}. The strengths of the various signals of the GJ\,3942 RVs are still too weak to make this claim, but it is however worth studying the most prominent of the periods resulting from the frequentist analysis at 10.4\,days. Its period ratio with GJ\,3942\,b at 6.9\,days is 3:2. This could be a strong sign for a 3:2 resonance which is, in contrast to the 3:1 resonance the 20.4-day signal would represent, a quite common configuration in multi-planetary systems as found by the Kepler results \citep{2011ApJS..197....8L, 2014ApJ...795...85W}. We show the chronological evolution of the periodogram power of the 10.4-day signal in Fig.\ref{f4ii}. We calculate the evolution both in the original data and also after subtracting the 16.3 and 6.9-day periods. As expected from a non-spurious signal, the significance improves with the treatment of the RV data. The signal came close to reaching the 0.1\% FAP level but the last 10 observations have somewhat degraded the significance. In the bottom panel of Fig.\,\ref{k2} we show the RV residuals of our MA method, after correcting for the periodic signals at 16.3 and 6.9\,days and phase-folded with a period of 10.38\,days. The resulting fit has a semi-amplitude of 2.52\,\ms. To address the dynamical stability of the putative system, we use the definitions given by \cite{2013MNRAS.436.3547G} and their Eq.\,3. As a result, we find that a two-planet system with orbital periods of 6.9 and 10.4\,days and masses of 7.2 and 6.3\,M$_{\oplus}$ can be be dynamically stable, albeit close to the limit and only for nearly circular orbits (eccentricity $<$ 0.2).

After considering the various scenarios, we can only confidently claim the existence of one planet around GJ\,3942, which we dub GJ\,3942\,b, and is likely to be a super-Earth with a minimum mass of 7.1\,M$_{\oplus}$ in a mildly-eccentric orbit with a period of 6.9\,days. Our best Keplerian fit of the periodicity with the MA method is shown on the bottom panel of Fig.\,\ref{k1}, the detailed characteristics are given in Table\,\ref{Tabresults}. To be able to estimate uncertainties of the fitted parameters of the Keplerian orbit (Eq.\,\ref{For2}), we use the IDL routine RVFIT \citep{2015PASP..127..567I} on the respective RV time-series data with the limits for the values as shown in the table. A further second planet in the system cannot be confirmed from our analysis of the available data and we thus deem it to be a candidate, with its possible properties also listed in Table\,\ref{Tabresults}. The rotational modulation signal is the strongest in the RV time-series data (and also in the activity indices and photometry) and is shown in Fig.\,\ref{k1} (top) and Table\,\ref{Tabresults} if modeled using the same parameters as Keplerian motion.

\section{Conclusions} \index{con} \label{con}

We investigated 145 spectroscopic observations of GJ\,3942 obtained over 3.3\,years with HARPS-N at the TNG in La Palma, Spain, and additional photometry from various sources, in particular the APACHE and EXORAP programs. We used RVs from the TERRA pipeline and activity indices calculated from H$\alpha$ and \ion{Ca}{II} emission lines originating in the stellar chromosphere. We investigate further the instrumental drifts of the observations and propose to add quadratically around 1\,\ms to the RV uncertainty if the second HARPS-N fiber is not used for simultaneous calibration (due to, e.g., long exposure times).

Three different approaches were used to search for planets. Significant in all of them is the one-planet model for which the moving average method delivers the lowest false alarm probability. Therefore, and since, in contrast to the multi-dimensional generalized Lomb-Scargle periodogram, it includes a treatment of the red noise, the results of this technique are the most reliable. The detailed Gaussian process regression, although probably best able to reproduce the changing characteristics of the magnetic periodicity, cannot resolve the GJ\,3942 system properly, since it adds the non-significant but existent 10.4-day signal into the noise or Gaussian process term. The one-planet model consists of two significant signals confirmed by all our considerations.

A dominant periodic signal at 16.28\,days is clearly visible in the photometric data, the activity indices, and the RV measurements. The RV data and the chromospheric indices show a phase shift of 120\,deg and we observe changes in the characteristics of the magnetic activity signal on rotational to seasonal timescales. We conclude that this period is closely related to the rotation of GJ\,3942 and the evolution of magnetically active regions on the stellar surface. Because of the complex nature, those signals cannot be described by single sinusoids. Together with possible resonances between the time-sampling and present periodicities, this introduces power at 8.15 and 5.4\,days, which are the first and second harmonics, respectively. It is interesting to note a rather sharp variation in the magnetic activity of the star at the end of the third season (August 2015), which seems also related with a brightness minimum in the photometric EXORAP data. Those variations can also be connected with the decreasing strength of the activity indices and the increasing RV scatter from seasons 1 to 4. The pattern changes also affect the detectability of further periodicities.

The second strongest signal in the periodogram at 6.91\,days has a high significance level that is not diminished by the modeling of magnetic activity. This periodic signal is not present in any activity index nor the photometric data and has a very stable period value. We therefore propose it to be induced by the orbit of a planet, GJ\,3942\,b. From our best fit we find a mass of 7.14$\pm$0.59\,M$_{\oplus}$ and an orbital semi-major axis of $a$=0.0608$\pm$0.0068\,AU (see Table\,\ref{Tabresults}). Using zero albedo, we calculate roughly an equilibrium temperature of 590\,K using $T_{\rm eq., p}=T_{\rm eff, *} \sqrt{R_{*}/2a}$. 

Further, we find still inconclusive evidence of at least one more planetary signal in our data, at 10.38\,days. Its significance rose steadily over the observational time-span and it is prominent in our analysis, but still lacks significance. The periodogram signal shows a visible harmonic at 5.2\,days and a yearly alias at 10.1\,days. This second planet, GJ\,3942\,c, would orbit with a 3:2 period ratio with GJ\,3942\,b and would have an equilibrium temperature of 515\,K. The possibility of additional planets in the system, likely in a near-resonant chain, is tantalizing but not confirmed by our analysis and thus remains open.

It is also interesting to assess the possibility of detecting transits from GJ\,3942\,b. We estimate the geometric transit probability at 4.7\% and a maximum duration of 2.5\,hours. From the stellar rotational period of 16.28$\pm$0.22\,days and a radius of 0.61$\pm$0.06~R$_{\odot}$ we calculate a rotational velocity of $v$=1.89$\pm$0.19~\kms. If we compare this with the $v \sin i$ measured from spectral features by \cite{2017A&A...598A..27M}, 1.67$\pm$0.30~\kms, we find it to be compatible with an inclination of the rotational axis of 90\,deg. Adopting the mass-radius relationship of exoplanets described in Fig.\,8 of \cite{2014ApJ...789..154D}, and assuming a 1.8\,R$_{\oplus}$ for a rocky configuration and 3\,R$_{\oplus}$ for a low-density planet, we calculate possible transit depths of 0.8 to 2.2\,mmag. This is challenging to detect with ground-based telescopes but could be an excellent target for the upcoming CHEOPS mission. The HADES observations on the other hand, are still ongoing in order to be able to find more hints for the existence of further planets around GJ\,3942, and to fulfill the prime goal of the program that is the discovery of a statistically significant sample of M-dwarf planets to provide stronger constraints on their occurrence rates, system architectures, and formation mechanisms.

\begin{acknowledgements}
	We are grateful to Guillem Anglada-Escud\'e for the HARPS-N TERRA code and the Likelihood-ratio diagrams. M. P., I. R., J. C. M., A. R., E. H, and M. L. acknowledge support from the Spanish Ministry of Economy and Competitiveness (MINECO) and the Fondo Europeo de Desarrollo Regional (FEDER) through grant ESP2016-80435-C2-1-R, as well as the support of the Generalitat de Catalunya/CERCA program. L.A. \& G.M. acknowledge support from the Ariel ASI-INAF agreement No. 2015-038-R.0. A.S.M., J.I.G.H., R.R.L. and B.T.P. acknowledge financial support from the Spanish Ministry project MINECO AYA2014-56359-P. J.I.G.H. also acknowledges financial support from the Spanish MINECO under the 2013 Ram\'on y Cajal program MINECO RYC-2013-14875. GAPS acknowledges support from INAF trough the "Progetti Premiali” funding scheme of the Italian Ministry of Education, University, and Research. The research leading to these results has received funding from the European Union Seventh Framework Programme (FP7/2007-2013) under Grant Agreement No. 313014 (ETAEARTH). The HARPS-N Project is a collaboration between the Astronomical Observatory of the Geneva University (lead), the CfA in Cambridge, the Universities of St. Andrews and Edinburgh, the Queen's University of Belfast, and the TNG-INAF Observatory. This paper makes use of data from the first public release of the WASP data as provided by the WASP consortium and services at the NASA Exoplanet Archive, which is operated by the California Institute of Technology, under contract with the National Aeronautics and Space Administration under the Exoplanet Exploration Program. This work has made further use of data from the European Space Agency (ESA) mission {\it Gaia} (\url{https://www.cosmos.esa.int/gaia}), processed by the {\it Gaia} Data Processing and Analysis Consortium (DPAC, url{https://www.cosmos.esa.int/web/gaia/dpac/consortium}). Funding for the DPAC has been provided by national institutions, in particular the institutions participating in the {\it Gaia} Multilateral Agreement.
\end{acknowledgements}

\bibliography{bibtex}{}
\bibliographystyle{aa}

\small
\onecolumn

\begin{longtable}{cccccccl}
	\caption{\label{Table} Data of 145 observations of GJ\,3942 including RV values of TERRA (drifts of second fiber already included) and DRS pipelines and their errors, the adopted instrumental drift, signal-to-noise ratios and the important activity indices.}\\

	\hline \hline
	\noalign{\smallskip}
	JD & RV TERRA & RV DRS   & inst. drift & S/N & $S$ index   & H$\alpha$ index   & Notes \\
	$[day]$-2\,456\,000 &   [m s$^{-1}$]  & [m s$^{-1}$] & [m s$^{-1}$] & order 46 &    & [10$^{-2}$]   & \\
	\hline
	\noalign{\smallskip}
	\endfirsthead
	\caption{continued.}\\

	\hline \hline
	\noalign{\smallskip}
	JD & RV TERRA & RV DRS   & inst. drift & S/N & $S$ index   & H$\alpha$ index   & Notes \\
	$[day]$-2\,456\,000 &   [m s$^{-1}$]  & [m s$^{-1}$] & [m s$^{-1}$] & order 46 &    & [10$^{-2}$]   & \\
	\hline
	\noalign{\smallskip}
	\endhead
	\hline
	\endfoot

	438.5877 & 0.613$\pm$1.113 & $-$18708.988$\pm$1.549 & 0.0$\pm$1.025 & 67.7 & 1.793$\pm$0.042 & 6.276$\pm$0.102 & (3) \\
	440.5944 & $-$8.092$\pm$0.812 & $-$18711.672$\pm$1.333 & 0.0$\pm$1.025 & 80.2 & 1.898$\pm$0.060 & 6.420$\pm$0.102 & (3) \\
	442.5249 & $-$3.829$\pm$0.797 & $-$18708.737$\pm$1.146 & 0.0$\pm$1.025 & 93.4 & 1.994$\pm$0.082 & 6.545$\pm$0.103 & (3) \\
	443.4686 & $-$6.004$\pm$1.134 & $-$18710.886$\pm$2.064 & 0.0$\pm$1.025 & 52.0 & 2.088$\pm$0.028 & 6.697$\pm$0.104 & (3) \\
	444.4712 & $-$6.527$\pm$0.941 & $-$18711.617$\pm$1.468 & 0.0$\pm$1.025 & 72.3 & 1.784$\pm$0.047 & 6.308$\pm$0.102 & (3) \\
	484.5573 & 6.111$\pm$1.028 & $-$18702.499$\pm$1.719 & $-$0.295$\pm$0.466 & 62.3 & 1.901$\pm$0.035 & 6.406$\pm$0.104 & (2) \\
	485.4829 & 4.955$\pm$0.803 & $-$18700.269$\pm$1.407 & 0.701$\pm$0.466 & 76.7 & 1.837$\pm$0.054 & 6.306$\pm$0.103 & (2) \\
	507.4481 & 5.466$\pm$2.473 & $-$18700.743$\pm$3.278 & 0.426$\pm$0.466 & 27.4 & 1.888$\pm$0.019 & 6.425$\pm$0.105 & (2) \\
	525.4515 & 1.035$\pm$1.841 & $-$18704.549$\pm$2.803 & 0.0$\pm$1.025 & 36.2 & 1.842$\pm$0.019 & 6.248$\pm$0.110 & (3) \\
	533.3811 & 9.353$\pm$0.842 & $-$18696.837$\pm$1.342 & 0.0$\pm$1.025 & 79.6 & 1.740$\pm$0.055 & 6.182$\pm$0.102 & (3) \\
	534.3513 & 6.998$\pm$1.171 & $-$18699.863$\pm$2.216 & 0.0$\pm$1.025 & 48.0 & 1.743$\pm$0.025 & 6.136$\pm$0.102 & (3) \\
	535.4050 & 0.740$\pm$1.098 & $-$18706.809$\pm$1.897 & 0.0$\pm$1.025 & 57.8 & 1.713$\pm$0.032 & 6.169$\pm$0.103 & (3) \\
	536.4488 & $-$2.641$\pm$1.018 & $-$18706.784$\pm$1.321 & 1.277$\pm$0.466 & 81.0 & 1.878$\pm$0.057 & 6.411$\pm$0.109 & (2) \\
	693.7262 & 0.464$\pm$1.207 & $-$18706.424$\pm$1.801 & 0.0$\pm$1.025 & 58.7 & 1.643$\pm$0.031 & 6.071$\pm$0.101 & (3) \\
	694.7261 & 4.463$\pm$1.199 & $-$18699.953$\pm$1.659 & 0.060$\pm$0.466 & 63.5 & 1.607$\pm$0.037 & 5.937$\pm$0.101 & (2) \\
	695.7478 & 7.208$\pm$1.636 & $-$18701.660$\pm$2.764 & 0.0$\pm$1.025 & 36.9 & 1.689$\pm$0.017 & 6.085$\pm$0.102 & (3) \\
	696.7101 & 4.767$\pm$0.811 & $-$18701.972$\pm$1.849 & 0.0$\pm$1.025 & 56.5 & 1.814$\pm$0.030 & 6.294$\pm$0.102 & (3) \\
	697.7231 & 5.559$\pm$0.928 & $-$18701.245$\pm$1.443 & 0.0$\pm$1.025 & 74.0 & 1.843$\pm$0.049 & 6.320$\pm$0.101 & (3) \\
	698.6999 & 2.548$\pm$1.291 & $-$18706.279$\pm$2.346 & 0.0$\pm$1.025 & 45.1 & 1.873$\pm$0.020 & 6.350$\pm$0.102 & (3) \\
	699.6711 & 0.834$\pm$0.985 & $-$18706.606$\pm$1.524 & 0.975$\pm$0.466 & 69.9 & 1.864$\pm$0.042 & 6.306$\pm$0.101 & (2) \\
	700.6779 & 0.340$\pm$0.874 & $-$18709.525$\pm$1.416 & 1.433$\pm$0.466 & 76.4 & 1.871$\pm$0.050 & 6.326$\pm$0.102 & (2) \\
	701.6547 & $-$1.396$\pm$1.232 & $-$18712.201$\pm$1.912 & 0.611$\pm$0.466 & 56.3 & 1.855$\pm$0.028 & 6.399$\pm$0.103 & (2) \\
	702.6503 & $-$6.852$\pm$1.096 & $-$18714.692$\pm$1.949 & $-$1.049$\pm$0.466 & 53.1 & 1.781$\pm$0.026 & 6.318$\pm$0.102 & (2) \\
	720.6468 & 0.062$\pm$0.919 & $-$18703.052$\pm$1.835 & 0.0$\pm$1.025 & 58.1 & 1.968$\pm$0.030 & 6.522$\pm$0.103 & (3) \\
	728.7639 & 0.579$\pm$1.272 & $-$18707.511$\pm$1.851 & 0.409$\pm$0.466 & 57.0 & 1.694$\pm$0.029 & 6.067$\pm$0.101 & (2) \\
	751.6805 & $-$3.484$\pm$1.078 & $-$18712.455$\pm$1.719 & 0.199$\pm$0.466 & 59.8 & 1.951$\pm$0.032 & 6.532$\pm$0.104 & (2) \\
	761.7075 & 14.241$\pm$1.443 & $-$18694.483$\pm$2.134 & $-$0.163$\pm$0.466 & 49.1 & 1.788$\pm$0.025 & 6.376$\pm$0.106 & (2) \\
	775.6407 & 3.451$\pm$1.295 & $-$18705.059$\pm$2.514 & 0.279$\pm$0.466 & 40.7 & 1.725$\pm$0.018 & 6.163$\pm$0.104 & (2) \\
	783.4505 & 0.549$\pm$0.969 & $-$18706.507$\pm$1.385 & 0.762$\pm$0.466 & 74.6 & 2.043$\pm$0.044 & 6.659$\pm$0.103 & (2) \\
	784.4391 & $-$8.868$\pm$0.831 & $-$18714.251$\pm$1.456 & $-$0.812$\pm$0.466 & 70.5 & 1.877$\pm$0.041 & 6.447$\pm$0.105 & (2) \\
	785.4442 & $-$12.351$\pm$1.365 & $-$18718.362$\pm$2.324 & $-$0.479$\pm$0.466 & 44.2 & 1.740$\pm$0.020 & 6.356$\pm$0.108 & (2) \\
	787.5590 & $-$6.514$\pm$0.996 & $-$18709.164$\pm$2.428 & $-$0.030$\pm$0.466 & 42.5 & 1.820$\pm$0.020 & 6.429$\pm$0.103 & (2) \\
	792.5216 & $-$4.871$\pm$1.207 & $-$18710.124$\pm$1.983 & $-$0.987$\pm$0.466 & 52.1 & 1.722$\pm$0.026 & 6.107$\pm$0.103 & (2) \\
	798.4424 & 5.681$\pm$1.013 & $-$18701.285$\pm$1.338 & $-$0.823$\pm$0.466 & 76.7 & 2.005$\pm$0.050 & 6.603$\pm$0.103 & (2) \\
	799.4656 & 3.747$\pm$0.875 & $-$18702.909$\pm$1.396 & 0.0$\pm$1.025 & 75.0 & 1.870$\pm$0.049 & 6.448$\pm$0.103 & (3) \\
	800.4462 & $-$1.510$\pm$0.983 & $-$18711.883$\pm$1.677 & 1.675$\pm$0.466 & 60.5 & 1.909$\pm$0.033 & 6.447$\pm$0.101 & (2) \\
	817.5506 & $-$1.367$\pm$0.805 & $-$18706.933$\pm$1.343 & $-$0.591$\pm$0.466 & 75.6 & 1.847$\pm$0.050 & 6.307$\pm$0.101 & (2) \\
	818.5510 & $-$3.388$\pm$0.854 & $-$18709.206$\pm$1.430 & 0.808$\pm$0.466 & 71.5 & 1.771$\pm$0.045 & 6.257$\pm$0.102 & (2) \\
	819.5843 & $-$8.134$\pm$1.030 & $-$18713.588$\pm$1.809 & 0.359$\pm$0.466 & 57.0 & 1.876$\pm$0.030 & 6.498$\pm$0.103 & (2) \\
	820.6290 & $-$2.488$\pm$1.150 & $-$18708.022$\pm$2.185 & 0.042$\pm$0.466 & 46.7 & 1.720$\pm$0.021 & 6.354$\pm$0.106 & (2) \\
	821.6261 & 0.401$\pm$1.146 & $-$18699.728$\pm$2.455 & 0.304$\pm$0.466 & 42.0 & 1.645$\pm$0.020 & 6.288$\pm$0.105 & (2) \\
	854.4557 & $-$5.136$\pm$0.935 & $-$18708.773$\pm$1.332 & 0.0$\pm$1.025 & 77.1 & 1.677$\pm$0.052 & 6.128$\pm$0.102 & (3) \\
	855.4438 & $-$4.291$\pm$0.904 & $-$18707.107$\pm$1.409 & 0.0$\pm$1.025 & 72.8 & 1.673$\pm$0.046 & 6.034$\pm$0.102 & (3) \\
	1029.7932 & $-$7.034$\pm$1.526 & $-$18715.058$\pm$2.530 & $-$0.404$\pm$0.466 & 39.8 & 1.816$\pm$0.018 & 6.349$\pm$0.103 & (2) \\
	1050.7481 & 3.703$\pm$2.247 & $-$18704.109$\pm$3.775 & 0.769$\pm$0.466 & 26.0 & 1.919$\pm$0.016 & 6.252$\pm$0.104 & (2) \\
	1090.7769 & 12.079$\pm$1.324 & $-$18690.562$\pm$2.438 & 0.0$\pm$1.025 & 38.5 & 1.790$\pm$0.018 & 6.249$\pm$0.102 & (3) \\
	1091.7112 & 12.841$\pm$1.080 & $-$18691.255$\pm$2.000 & 0.0$\pm$1.025 & 48.5 & 1.777$\pm$0.023 & 6.250$\pm$0.102 & (3) \\
	1092.7233 & 10.537$\pm$0.875 & $-$18694.775$\pm$1.743 & 0.0$\pm$1.025 & 55.6 & 2.020$\pm$0.029 & 6.548$\pm$0.102 & (3) \\
	1093.7118 & 2.184$\pm$0.914 & $-$18701.860$\pm$1.626 & 0.0$\pm$1.025 & 61.9 & 1.819$\pm$0.035 & 6.265$\pm$0.102 & (3) \\
	1094.7174 & $-$5.480$\pm$1.192 & $-$18710.880$\pm$1.851 & 0.0$\pm$1.025 & 50.6 & 1.928$\pm$0.026 & 6.353$\pm$0.102 & (3) \\
	1096.7402 & $-$4.880$\pm$1.538 & $-$18709.660$\pm$2.809 & 0.0$\pm$1.025 & 35.0 & 1.759$\pm$0.016 & 6.277$\pm$0.103 & (3) \\
	1097.7125 & $-$3.481$\pm$0.856 & $-$18707.293$\pm$1.654 & 0.082$\pm$0.466 & 59.5 & 1.734$\pm$0.032 & 6.165$\pm$0.102 & (2) \\
	1098.7120 & $-$2.737$\pm$0.964 & $-$18708.411$\pm$1.748 & 0.0$\pm$1.025 & 56.5 & 1.665$\pm$0.031 & 6.094$\pm$0.102 & (3) \\
	1099.7075 & $-$0.551$\pm$1.150 & $-$18703.119$\pm$2.117 & 0.516$\pm$0.466 & 46.5 & 1.656$\pm$0.022 & 6.089$\pm$0.103 & (2) \\
	1113.5836 & $-$2.870$\pm$1.182 & $-$18708.141$\pm$1.849 & 0.496$\pm$0.466 & 54.7 & 1.664$\pm$0.027 & 6.108$\pm$0.102 & (2) \\
	1114.5069 & -- & $-$18706.594$\pm$5.639 & 0.741$\pm$0.466 & 18.3 & 1.288$\pm$0.017 & 6.043$\pm$0.111 & (1),(4) \\
	1114.6085 & -- & $-$18719.875$\pm$10.021 & 0.0$\pm$1.025 & 11.0 & 1.222$\pm$0.025 & 6.008$\pm$0.121 & (3),(4) \\
	1115.5025 & $-$2.979$\pm$1.103 & $-$18705.707$\pm$2.072 & 0.100$\pm$0.466 & 48.3 & 1.566$\pm$0.021 & 5.976$\pm$0.104 & (2) \\
	1116.5643 & $-$5.335$\pm$0.788 & $-$18710.515$\pm$1.385 & 0.127$\pm$0.466 & 72.4 & 1.658$\pm$0.046 & 6.040$\pm$0.102 & (2) \\
	1117.6055 & $-$4.947$\pm$1.307 & $-$18712.244$\pm$1.980 & 0.0$\pm$1.025 & 50.7 & 1.720$\pm$0.025 & 6.122$\pm$0.104 & (3) \\
	1137.7257 & $-$4.772$\pm$0.986 & $-$18708.514$\pm$1.712 & $-$0.064$\pm$0.466 & 55.4 & 1.670$\pm$0.030 & 6.016$\pm$0.102 & (2) \\
	1139.6486 & 9.596$\pm$0.946 & $-$18694.519$\pm$1.552 & 1.261$\pm$0.466 & 64.0 & 1.717$\pm$0.037 & 6.111$\pm$0.102 & (2) \\
	1142.6156 & 1.357$\pm$0.973 & $-$18704.958$\pm$1.591 & 0.381$\pm$0.466 & 64.5 & 1.829$\pm$0.040 & 6.385$\pm$0.103 & (2) \\
	1143.5760 & $-$5.004$\pm$0.856 & $-$18711.518$\pm$1.555 & 0.097$\pm$0.466 & 65.8 & 1.868$\pm$0.040 & 6.438$\pm$0.104 & (2) \\
	1144.5698 & $-$5.182$\pm$1.012 & $-$18710.995$\pm$1.787 & 0.121$\pm$0.466 & 57.2 & 1.852$\pm$0.031 & 6.439$\pm$0.103 & (2) \\
	1146.6834 & $-$0.186$\pm$1.107 & $-$18705.001$\pm$1.703 & 1.000$\pm$0.466 & 57.3 & 1.798$\pm$0.031 & 6.271$\pm$0.106 & (2) \\
	1147.6914 & 2.240$\pm$1.067 & $-$18697.785$\pm$1.679 & 1.120$\pm$0.466 & 59.0 & 1.696$\pm$0.032 & 6.070$\pm$0.104 & (2) \\
	1148.6048 & $-$0.233$\pm$0.742 & $-$18705.134$\pm$1.436 & 0.0$\pm$1.025 & 71.9 & 1.690$\pm$0.047 & 6.144$\pm$0.103 & (3) \\
	1159.5110 & $-$3.910$\pm$0.888 & $-$18712.010$\pm$1.707 & 0.623$\pm$0.466 & 61.1 & 2.026$\pm$0.035 & 6.601$\pm$0.104 & (2) \\
	1166.5414 & 5.272$\pm$0.790 & $-$18703.092$\pm$1.481 & 0.0$\pm$1.025 & 67.6 & 1.680$\pm$0.043 & 6.066$\pm$0.102 & (3) \\
	1170.5532 & 2.897$\pm$0.713 & $-$18703.727$\pm$1.277 & 0.155$\pm$0.466 & 79.1 & 1.858$\pm$0.057 & 6.285$\pm$0.102 & (2) \\
	1172.6577 & 2.984$\pm$0.875 & $-$18700.070$\pm$1.607 & $-$0.722$\pm$0.466 & 62.0 & 1.925$\pm$0.033 & 6.338$\pm$0.102 & (2) \\
	1173.5440 & 1.916$\pm$0.932 & $-$18700.902$\pm$1.772 & $-$0.078$\pm$0.466 & 56.2 & 1.853$\pm$0.028 & 6.345$\pm$0.102 & (2) \\
	1174.4194 & 3.283$\pm$1.422 & $-$18703.970$\pm$2.337 & 0.728$\pm$0.466 & 43.3 & 1.824$\pm$0.020 & 6.429$\pm$0.103 & (2) \\
	1175.5652 & 1.982$\pm$1.304 & $-$18701.155$\pm$2.208 & 1.929$\pm$0.466 & 44.8 & 1.855$\pm$0.020 & 6.367$\pm$0.103 & (2) \\
	1176.5583 & $-$3.793$\pm$1.421 & $-$18704.542$\pm$2.491 & 1.046$\pm$0.466 & 40.1 & 1.908$\pm$0.018 & 6.498$\pm$0.105 & (2) \\
	1177.5052 & $-$6.194$\pm$1.052 & $-$18713.860$\pm$1.678 & 0.263$\pm$0.466 & 60.8 & 1.928$\pm$0.033 & 6.457$\pm$0.103 & (2) \\
	1178.4935 & $-$11.795$\pm$1.723 & $-$18717.584$\pm$2.861 & 0.284$\pm$0.466 & 34.9 & 1.699$\pm$0.015 & 6.369$\pm$0.105 & (2) \\
	1203.4710 & 8.736$\pm$0.738 & $-$18692.305$\pm$1.351 & 0.697$\pm$0.466 & 73.7 & 1.831$\pm$0.047 & 6.315$\pm$0.102 & (2) \\
	1204.4918 & 8.409$\pm$0.867 & $-$18696.524$\pm$1.378 & 0.870$\pm$0.466 & 72.6 & 2.003$\pm$0.047 & 6.533$\pm$0.102 & (2) \\
	1209.5459 & 4.999$\pm$1.338 & $-$18699.244$\pm$2.279 & 1.809$\pm$0.466 & 44.1 & 1.884$\pm$0.020 & 6.488$\pm$0.103 & (2) \\
	1211.6177 & $-$2.961$\pm$2.534 & $-$18704.829$\pm$3.601 & 0.0$\pm$1.025 & 27.1 & 1.811$\pm$0.014 & 6.381$\pm$0.111 & (3) \\
	1239.3972 & $-$3.077$\pm$0.713 & $-$18703.594$\pm$1.404 & 1.132$\pm$0.068 & 72.7 & 1.900$\pm$0.047 & 6.316$\pm$0.102 & (1) \\
	1240.3978 & $-$5.618$\pm$0.773 & $-$18708.907$\pm$1.462 & $-$0.214$\pm$0.068 & 69.8 & 1.839$\pm$0.043 & 6.246$\pm$0.104 & (1) \\
	1241.3944 & $-$6.450$\pm$0.860 & $-$18707.593$\pm$1.893 & 0.722$\pm$0.068 & 53.2 & 1.863$\pm$0.027 & 6.266$\pm$0.103 & (1) \\
	1242.3944 & $-$5.128$\pm$1.116 & $-$18706.990$\pm$1.953 & 0.927$\pm$0.068 & 52.4 & 1.899$\pm$0.027 & 6.329$\pm$0.104 & (1) \\
	1249.4140 & 0.505$\pm$1.290 & $-$18702.351$\pm$2.332 & 1.075$\pm$0.068 & 43.9 & 1.753$\pm$0.022 & 6.178$\pm$0.109 & (1) \\
	1250.4054 & 0.651$\pm$0.885 & $-$18702.769$\pm$1.460 & $-$0.053$\pm$0.069 & 69.4 & 1.821$\pm$0.043 & 6.199$\pm$0.103 & (1) \\
	1251.3822 & 2.025$\pm$0.753 & $-$18700.268$\pm$1.378 & 0.289$\pm$0.069 & 73.7 & 1.802$\pm$0.049 & 6.204$\pm$0.103 & (1) \\
	1260.3813 & $-$8.159$\pm$1.053 & $-$18712.230$\pm$1.284 & 0.095$\pm$0.070 & 80.8 & 1.845$\pm$0.061 & 6.360$\pm$0.104 & (1) \\
	1261.3670 & $-$8.724$\pm$0.695 & $-$18714.146$\pm$1.334 & 0.259$\pm$0.069 & 77.4 & 1.751$\pm$0.055 & 6.193$\pm$0.102 & (1) \\
	1262.3632 & $-$4.863$\pm$0.907 & $-$18714.465$\pm$1.483 & $-$0.845$\pm$0.070 & 69.5 & 1.715$\pm$0.046 & 6.110$\pm$0.102 & (1) \\
	1263.3642 & 3.861$\pm$0.811 & $-$18702.245$\pm$1.355 & $-$1.883$\pm$0.070 & 75.4 & 1.774$\pm$0.053 & 6.215$\pm$0.103 & (1) \\
	1264.3629 & 10.473$\pm$1.169 & $-$18692.545$\pm$1.945 & 0.180$\pm$0.070 & 52.4 & 1.739$\pm$0.027 & 6.176$\pm$0.106 & (1) \\
	1283.3710 & $-$1.263$\pm$0.909 & $-$18708.972$\pm$1.479 & 0.113$\pm$0.108 & 69.8 & 1.768$\pm$0.045 & 6.229$\pm$0.104 & (1) \\
	1306.3299 & 1.699$\pm$0.913 & $-$18703.576$\pm$1.668 & 0.462$\pm$0.085 & 62.1 & 1.862$\pm$0.035 & 6.343$\pm$0.106 & (1) \\
	1307.3250 & 3.006$\pm$0.845 & $-$18702.064$\pm$1.350 & 0.226$\pm$0.085 & 76.8 & 1.783$\pm$0.054 & 6.220$\pm$0.108 & (1) \\
	1308.3250 & 0.551$\pm$0.780 & $-$18700.805$\pm$1.235 & 1.783$\pm$0.086 & 84.8 & 1.782$\pm$0.062 & 6.204$\pm$0.107 & (1) \\
	1443.7135 & 1.047$\pm$1.255 & $-$18704.036$\pm$2.864 & 0.0$\pm$1.025 & 34.1 & 1.623$\pm$0.017 & 6.143$\pm$0.103 & (3) \\
	1444.6979 & 5.268$\pm$1.259 & $-$18696.666$\pm$2.387 & 0.0$\pm$1.025 & 41.6 & 1.638$\pm$0.021 & 6.130$\pm$0.103 & (3) \\
	1445.7050 & 10.455$\pm$1.519 & $-$18695.824$\pm$2.907 & 0.0$\pm$1.025 & 34.2 & 1.653$\pm$0.017 & 6.219$\pm$0.103 & (3) \\
	1472.6496 & $-$1.920$\pm$1.179 & $-$18706.845$\pm$2.259 & 0.0$\pm$1.025 & 44.4 & 1.626$\pm$0.023 & 6.020$\pm$0.103 & (3) \\
	1474.6541 & 0.754$\pm$1.152 & $-$18700.737$\pm$1.866 & 0.0$\pm$1.025 & 54.1 & 1.541$\pm$0.030 & 5.902$\pm$0.104 & (3) \\
	1475.6334 & 2.257$\pm$1.302 & $-$18705.561$\pm$2.287 & 0.0$\pm$1.025 & 44.0 & 1.573$\pm$0.023 & 6.031$\pm$0.102 & (3) \\
	1491.5313 & 5.065$\pm$0.951 & $-$18697.514$\pm$1.662 & 0.0$\pm$1.025 & 58.4 & 1.599$\pm$0.033 & 5.988$\pm$0.103 & (3) \\
	1492.7096 & 9.621$\pm$0.887 & $-$18694.305$\pm$1.550 & 0.0$\pm$1.025 & 64.0 & 1.675$\pm$0.040 & 6.134$\pm$0.103 & (3) \\
	1501.6133 & $-$6.073$\pm$1.170 & $-$18712.781$\pm$1.908 & 0.0$\pm$1.025 & 51.8 & 1.717$\pm$0.027 & 6.253$\pm$0.103 & (3) \\
	1502.7151 & $-$12.367$\pm$1.008 & $-$18715.081$\pm$1.596 & 0.0$\pm$1.025 & 62.1 & 1.738$\pm$0.037 & 6.262$\pm$0.105 & (3) \\
	1508.5400 & 8.009$\pm$1.075 & $-$18695.271$\pm$1.636 & $-$0.173$\pm$0.115 & 57.2 & 1.671$\pm$0.033 & 6.063$\pm$0.103 & (1) \\
	1509.5383 & 1.406$\pm$1.502 & $-$18698.684$\pm$2.465 & 1.863$\pm$0.117 & 36.1 & 1.849$\pm$0.018 & 6.265$\pm$0.102 & (1) \\
	1510.5188 & 7.906$\pm$1.248 & $-$18694.940$\pm$2.142 & 0.457$\pm$0.116 & 41.3 & 1.793$\pm$0.021 & 6.106$\pm$0.102 & (1) \\
	1513.6474 & 2.332$\pm$1.212 & $-$18703.631$\pm$1.943 & 0.0$\pm$1.025 & 51.5 & 1.784$\pm$0.027 & 6.292$\pm$0.103 & (3) \\
	1523.5068 & $-$4.022$\pm$1.253 & $-$18709.428$\pm$2.388 & 0.0$\pm$1.025 & 41.5 & 1.585$\pm$0.020 & 6.086$\pm$0.105 & (3) \\
	1524.5078 & $-$2.606$\pm$1.407 & $-$18706.911$\pm$2.623 & 0.0$\pm$1.025 & 37.5 & 1.654$\pm$0.017 & 6.049$\pm$0.104 & (3) \\
	1525.5334 & $-$0.837$\pm$1.292 & $-$18705.145$\pm$2.143 & 0.0$\pm$1.025 & 46.2 & 1.653$\pm$0.023 & 6.145$\pm$0.103 & (3) \\
	1526.5191 & 7.131$\pm$1.019 & $-$18695.606$\pm$1.924 & 0.0$\pm$1.025 & 52.3 & 1.785$\pm$0.029 & 6.285$\pm$0.103 & (3) \\
	1537.4831 & $-$3.482$\pm$1.150 & $-$18706.644$\pm$2.399 & 0.070$\pm$0.067 & 41.7 & 1.763$\pm$0.021 & 6.340$\pm$0.103 & (1) \\
	1538.5009 & $-$0.371$\pm$0.934 & $-$18699.647$\pm$1.551 & 0.179$\pm$0.067 & 62.5 & 1.727$\pm$0.039 & 6.206$\pm$0.102 & (1) \\
	1540.6220 & 4.347$\pm$1.215 & $-$18698.685$\pm$2.446 & $-$0.806$\pm$0.067 & 40.2 & 1.772$\pm$0.020 & 6.188$\pm$0.103 & (1) \\
	1549.6026 & 5.670$\pm$1.480 & $-$18697.535$\pm$2.367 & $-$0.850$\pm$0.066 & 35.7 & 1.896$\pm$0.018 & 6.402$\pm$0.103 & (1) \\
	1550.6090 & $-$5.436$\pm$1.181 & $-$18707.630$\pm$1.701 & 0.668$\pm$0.066 & 51.8 & 1.785$\pm$0.029 & 6.228$\pm$0.103 & (1) \\
	1551.5963 & $-$16.538$\pm$1.479 & $-$18721.479$\pm$2.590 & 0.258$\pm$0.066 & 36.9 & 1.729$\pm$0.018 & 6.333$\pm$0.103 & (1) \\
	1552.5768 & $-$16.093$\pm$0.915 & $-$18719.165$\pm$1.379 & $-$0.125$\pm$0.067 & 72.1 & 1.674$\pm$0.049 & 6.129$\pm$0.103 & (1) \\
	1553.5733 & $-$13.750$\pm$1.477 & $-$18714.503$\pm$2.365 & 0.393$\pm$0.067 & 41.7 & 1.653$\pm$0.021 & 6.159$\pm$0.102 & (1) \\
	1594.4763 & $-$1.909$\pm$1.036 & $-$18703.278$\pm$1.993 & 0.0$\pm$1.025 & 48.8 & 1.892$\pm$0.026 & 6.380$\pm$0.105 & (3) \\
	1596.4883 & 7.675$\pm$1.179 & $-$18700.045$\pm$2.174 & 0.0$\pm$1.025 & 46.9 & 1.866$\pm$0.025 & 6.522$\pm$0.107 & (3) \\
	1597.4821 & 7.769$\pm$1.415 & $-$18697.408$\pm$2.067 & 0.0$\pm$1.025 & 46.8 & 1.980$\pm$0.024 & 6.519$\pm$0.104 & (3) \\
	1603.4676 & $-$1.844$\pm$0.989 & $-$18704.734$\pm$1.582 & 0.0$\pm$1.025 & 59.9 & 1.551$\pm$0.036 & 5.952$\pm$0.102 & (3) \\
	1604.4582 & $-$2.862$\pm$1.265 & $-$18703.700$\pm$1.993 & 0.0$\pm$1.025 & 49.4 & 1.558$\pm$0.026 & 6.100$\pm$0.104 & (3) \\
	1606.3684 & $-$0.496$\pm$1.151 & $-$18698.794$\pm$1.992 & $-$0.065$\pm$0.073 & 49.7 & 1.670$\pm$0.027 & 6.030$\pm$0.105 & (1) \\
	1607.4869 & 9.152$\pm$1.704 & $-$18689.137$\pm$2.676 & $-$0.427$\pm$0.073 & 33.2 & 1.677$\pm$0.018 & 6.117$\pm$0.105 & (1) \\
	1608.3691 & 10.899$\pm$0.914 & $-$18687.570$\pm$1.427 & 0.248$\pm$0.073 & 70.7 & 1.888$\pm$0.050 & 6.444$\pm$0.103 & (1) \\
	1609.3773 & 8.120$\pm$1.719 & $-$18693.625$\pm$3.262 & 0.385$\pm$0.073 & 30.8 & 1.897$\pm$0.019 & 6.392$\pm$0.104 & (1) \\
	1610.3713 & $-$0.940$\pm$2.180 & $-$18705.357$\pm$4.270 & 0.430$\pm$0.073 & 23.9 & 1.701$\pm$0.018 & 6.342$\pm$0.109 & (1) \\
	1620.4167 & $-$4.784$\pm$1.444 & $-$18705.613$\pm$2.108 & 0.0$\pm$1.025 & 47.0 & 1.487$\pm$0.024 & 5.934$\pm$0.104 & (3) \\
	1621.4386 & -- & $-$18697.552$\pm$5.357 & 0.0$\pm$1.025 & 18.1 & 1.211$\pm$0.019 & 5.911$\pm$0.110 & (3),(4) \\
	1625.3929 & 7.426$\pm$0.979 & $-$18694.095$\pm$1.442 & 0.162$\pm$0.073 & 71.1 & 1.855$\pm$0.050 & 6.405$\pm$0.102 & (1) \\
	1626.3960 & 0.686$\pm$1.359 & $-$18702.835$\pm$2.026 & $-$0.163$\pm$0.074 & 50.8 & 1.774$\pm$0.027 & 6.256$\pm$0.102 & (1) \\
	1627.3906 & $-$2.227$\pm$1.102 & $-$18708.008$\pm$1.803 & $-$0.948$\pm$0.073 & 57.3 & 1.782$\pm$0.035 & 6.383$\pm$0.104 & (1) \\
	1628.3937 & $-$1.839$\pm$1.098 & $-$18702.884$\pm$2.012 & $-$0.223$\pm$0.074 & 50.0 & 1.717$\pm$0.028 & 6.303$\pm$0.111 & (1) \\
	1629.3893 & 5.080$\pm$1.172 & $-$18699.319$\pm$1.628 & 0.150$\pm$0.074 & 63.6 & 1.778$\pm$0.041 & 6.310$\pm$0.106 & (1) \\
	1630.3921 & $-$2.377$\pm$1.282 & $-$18711.553$\pm$2.206 & 0.162$\pm$0.074 & 46.7 & 1.804$\pm$0.025 & 6.405$\pm$0.110 & (1) \\
	1632.3916 & $-$12.258$\pm$1.120 & $-$18716.701$\pm$1.601 & 0.180$\pm$0.074 & 63.4 & 1.786$\pm$0.041 & 6.386$\pm$0.103 & (1) \\
	1638.3967 & 2.587$\pm$1.081 & $-$18703.416$\pm$1.758 & $-$1.298$\pm$0.074 & 57.5 & 1.529$\pm$0.035 & 6.005$\pm$0.102 & (1) \\
	1642.3937 & 1.767$\pm$1.152 & $-$18700.358$\pm$1.719 & $-$0.065$\pm$0.074 & 59.7 & 1.740$\pm$0.037 & 6.326$\pm$0.103 & (1) \\
	\noalign{\smallskip}
	\hline \hline

\end{longtable}
\tablefoot{{\tablefoottext{1}{instrumental drift measured with second fiber}}, \tablefoottext{2}{instrumental drift calculated (including additional error of 0.466\,\ms)}, \tablefoottext{3}{no instrumental drift available (including additional error of 1.025\,\ms)}, \tablefoottext{4}{spectra not used due to low S/N and high RV error.}}
\end{document}